\documentclass[aps,groupedaddress,superscriptaddress,amsmath,amssymb,twocolumn,prb]{revtex4-2}
\usepackage{bm}
\usepackage[pdftex]{graphicx}
\usepackage{xcolor}
\usepackage{color}
\usepackage[normalem]{ulem}
\usepackage{enumerate}
\usepackage{epstopdf}
\usepackage{braket}
\usepackage{hyperref}
\usepackage{lineno}
\usepackage{blindtext}
\usepackage{comment}
\usepackage{amsmath}

\begin{document}

\title{Chern junctions in Moiré-Patterned Graphene/$\rm PbI_2$}

\author{Yan Sun}
\affiliation{Laboratoire de Physique des Solides, Universit\'e Paris-Saclay, CNRS, 91405 Orsay, France}
\author{M. Monteverde}
\affiliation{Laboratoire de Physique des Solides, Universit\'e Paris-Saclay, CNRS, 91405 Orsay, France}
\author{V Derkach}
\affiliation{O. Ya. Usikov Institute for Radiophysics and Electronics, National Academy of Sciences of Ukraine, 12 Academician Proskury Street, Kharkov 61085, Ukraine}
\author{K. Watanabe}
\affiliation{Research Center for Electronic and Optical Materials, National Institute for Materials Science, 1-1 Namiki, Tsukuba, 305-0044, Japan}
\author{T. Taniguchi}
\affiliation{International Center for Materials Nanoarchitectonics, National Institute for Materials Science, 1-1 Namiki, Tsukuba, 305-0044, Japan}
\author{F. Chiodi}
\affiliation{Centre de nanosciences et de nanotechnologies, Universit\'e Paris-Saclay, CNRS, 91120 Palaiseau, France}
\author{H. Bouchiat}
\affiliation{Laboratoire de Physique des Solides, Universit\'e Paris-Saclay, CNRS, 91405 Orsay, France}
\author{A.D. Chepelianskii}
\affiliation{Laboratoire de Physique des Solides, Universit\'e Paris-Saclay, CNRS, 91405 Orsay, France}

\begin{abstract}

  Expanding the moiré material library continues to reveal novel quantum phases and emergent electronic behaviors. Here, we introduce $\mathrm{PbI}_2$ into the moiré family and investigate the magnetotransport properties of a hexagonal boron nitride/graphene/$\mathrm{PbI}_2$ heterostructure. In the high-field quantum Hall regime, we observe robust dissipationless transport at the charge neutrality point, indicative of incompressible states at filling factor $\nu_h = 0$.
  A fractional conductance plateau at $2/3,e^2/h$ emerges, which we attribute to a Chern junction formed between domains with distinct Chern numbers arising from moiré-modulated and conventional quantum Hall states. Additionally, coherent electronic interference appears along trajectories associated with the $\nu_m = -2$ Chern state. These observations provide compelling evidence for the formation of moiré domains that nontrivially interrupt incompressible quantum Hall states, reflecting the strong moiré potential in the BN/graphene/$\mathrm{PbI}_2$ superlattice. We suggest that the moiré Hofstadter spectrum coupled with the proximity-induced spin–orbit interaction from $\mathrm{PbI}_2$ gives rise to a high magnetic field topological insulator phase explaining ballistic transport at the charge neutrality point in the graphene monolayer. 
\end{abstract}

\maketitle 
\section{INTRODUCTION}
The discovery of correlated electronic phases in moiré superlattices has revolutionized our understanding of quantum matter in two-dimensional (2D) systems. By introducing a twist or lattice mismatch between atomically thin crystals, moiré engineering enables the creation of highly tunable quantum systems with emergent properties governed by electronic correlations, topology, and band reconstruction \cite{cao2018unconventional,andrei2020graphene,ribeiro2018twistable}. These systems provide unprecedented opportunities to investigate strongly interacting quantum states, including Mott insulators, superconductivity, topological flat bands, and fractional quantum Hall (QH) effects \cite{dean2013hofstadter,xia2025superconductivity,nuckolls2024microscopic}. 

A growing number of moiré materials - such as twisted bilayer graphene, graphene/hexagonal boron nitride (BN) heterostructures, and transition metal dichalcogenide bilayers (TMDs) - have demonstrated that delicate control of interlayer coupling, twist angle, and dielectric environment can lead to rich and tunable phase diagrams. Recent efforts have focused on expanding the moiré material library by incorporating components with intrinsic magnetic order or strong spin–orbit coupling, as well as the development of higher-order superlattice architectures, such as moiré-of-moiré structures, aiming to further enrich the quantum behavior accessible in these heterostructures \cite{rothstein2024band,mak2019probing,li2021imaging,mak2022semiconductor,lai2025moire,wang2025moire,xie2025strong}.

In this context, the incorporation of lead iodide ($\mathrm{PbI}_2$), a layered semiconductor with strong intrinsic spin–orbit coupling, opens new possibilities for designing moiré systems with enhanced quantum effects. 
$\mathrm{PbI}_2$ possesses a wide bandgap ranging from 2.1 eV to 2.6 eV depending on its thickness, making it widely used in photon detector, for instance in X-ray detectors \cite{yagmurcukardes2018electronic, zhang2018low}. $\rm PbI_2$ is also a precursor for metalic halide perovskites, which display strong charge transfer with graphene depending on the interfacial termination of the perovskite layer \cite{sun2024quantum,zhong2024growth}.  
Its ability to form various van der Waals heterojunctions via band alignment engineering with TMDs has been reported \cite{sun2020engineering}. While $\mathrm{PbI}_2$ has been extensively studied in optoelectronic applications, its integration into quantum transport platforms remains largely unexplored.

Crystallographically, $\mathrm{PbI}_2$ adopts a layered honeycomb structure with a lattice constant of a = 4.59 $\rm \AA$, approximately twice that of monolayer graphene (2.46 $\rm \AA$), favoring the formation of moiré superlattices in graphene/$\mathrm{PbI}_2$ heterostructures, a phenomenon that has been visualized using scanning tunneling microscopy \cite{sinha2020atomic, wang2024crystal}. 
The presence of heavy atoms gives rise to strong spin–orbit coupling (SOC), that can be induced in graphene by proximity effect, a key ingredient for realizing topological insulating phases and quantum spin Hall states \cite{wang2015strong, wakamura2018strong, island2019spin}.   Theoretical proposals have predicted quantum anomalous Hall states in stanene/$\mathrm{PbI}_2$ heterostructures, further highlighting the potential of $\mathrm{PbI}_2$ for exploring spintronic phenomena and quantum spin Hall effects \cite{zhang2016quantum, zhang2017room}.

\begin{figure*}[ht]
\centerline{
\includegraphics[clip=true,width=0.8\textwidth]{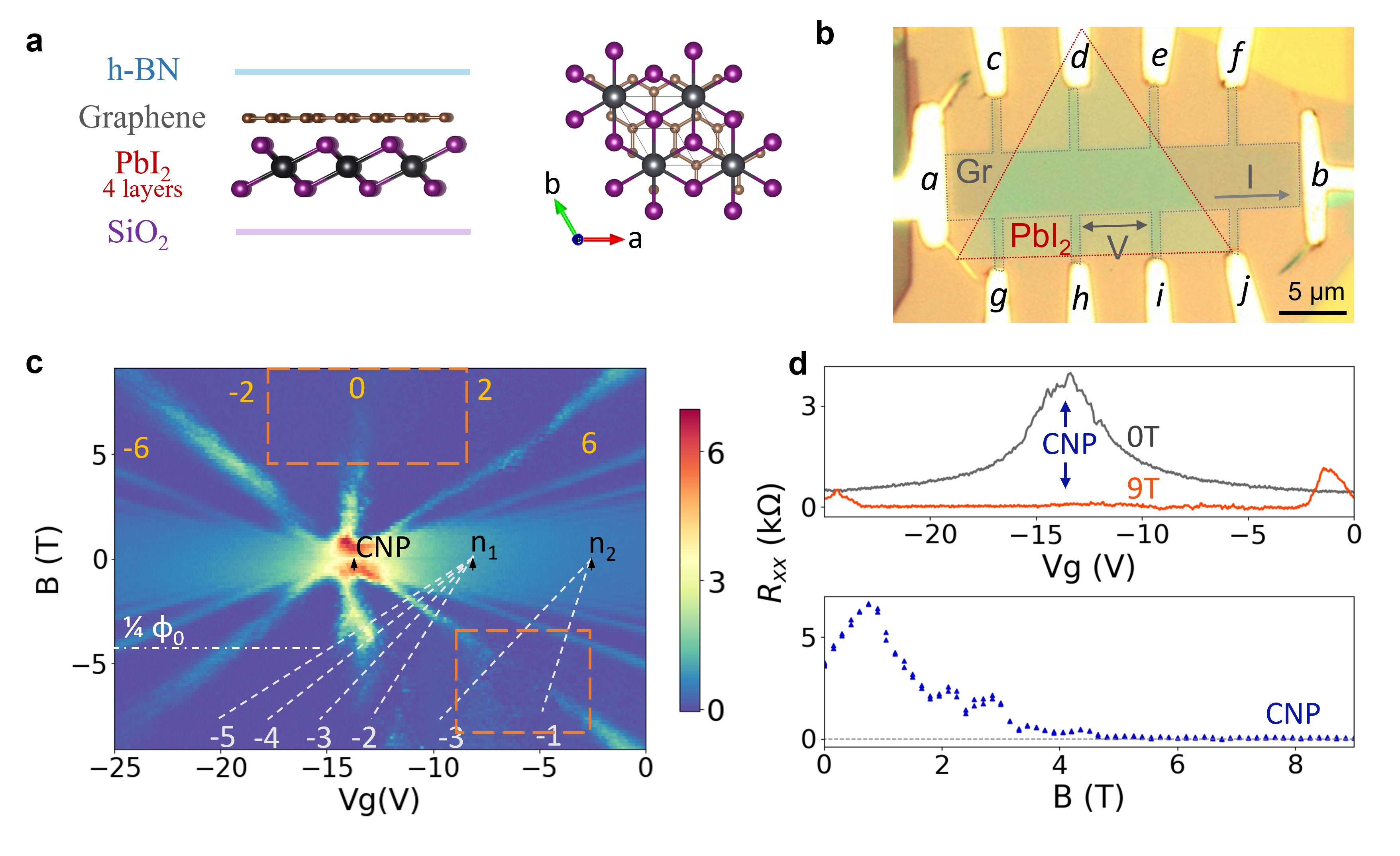} 
}
\caption{a) Schematic cross-section of the architecture (left), alongside bottom view of the $\rm PbI_2$ (1T phase)/Graphene lattice structure. b) Optical image of the sample with letters labeling the electrodes. c) Landau fan diagram of longitudinal resistance $\rm R_{h-i}$ as a function of gate voltage and magnetic field, measured via four-probe configuration. Yellow numerical labels indicate Landau level filling factors corresponding to resistance minima in the quantum Hall regime. The orange square highlights the suppression of the resistance peak of $\rm R_{xx}$ with increasing magnetic field, signaling a robust  $\rm R_{xx}$=0 state. Dashed white lines show the link between the onset of the vanishing $\rm R_{xx}$ and the moiré structures decribed later. d) $\rm R_{xx}$ as function of gate voltage at 0 T and 9 T (upper panel), a slice of the magnetic-field dependence at the charge neutrality point (lower panel), showing that resistance vanishes at B $>$ 4.5 T.} 
\label{sample}
\end{figure*}

Here, we fabricate BN/graphene/$\mathrm{PbI}_2$ heterostructures and investigate the resulting moiré superlattice by low-temperature high-field magnetotransport measurements. We observe a vanishing longitudinal resistance ($\rm R_{xx} = 0$) at the charge neutrality point (CNP) in perpendicular magnetic fields ($B > 4.5\ \mathrm{T}$), indicating ballistic transport mediated by moiré-induced channels. A robust $2/3$ conductance plateau emerges on the electron-doped side, suggesting the formation of unconventional correlated states. Both the $2/3$ plateau and the $\rm  R_{xx} = 0$ state are bounded by linear trajectories associated with moiré band filling indices $s_m = 1$ and $s_m = 0$, consistent with junctions between integer $\nu_h = 2$ QH states and moiré $\nu_m = -2$ domains. Moreover, we identify three distinct families of incompressible states in two-probe resistance measurements, reflecting three different electronic flavors in the moiré minibands. These are likely influenced by spin–orbit interaction induced by the $\mathrm{PbI}_2$ layer, as evidenced by quantum beating patterns observed in Shubnikov–de Haas oscillations. 
The marked modifications in magnetotransport at CNP suggests that $\mathrm{PbI}_2$ induces a significant moiré potential on graphene, and the moiré-induced effects become evident in high magnetic fields, manifesting as coherent resistance fluctuation and electronic interference patterns.

The Kane-Mele model seems the only known theoretical framework that can explain ballistic conduction channels at the Dirac point in graphene, both at zero magnetic field and in the presence of strong magnetic fields. However, inducing a sufficiently strong Kane–Mele spin–orbit interaction in graphene to realize a topological insulator state has so far remained elusive. In heterostructures with transition-metal dichalcogenides, other types of spin-orbit interactions-such as Rashba and valley-Zeeman couplings are present competing with the Kane-Mele term \cite{wang2015strong,wakamura2018strong,rao2023ballistic}.
We have show that a strong spin-orbit interaction is induced by the $\mathrm{PbI}_2$ layer; however, the precise nature of the induced coupling is not yet fully understood. It is possible that the interplay between moiré subbands and spin-orbit interactions \cite{li2019twist} transforms certain moiré domains of our heterostructure into a topological insulator at high magnetic fields. Such a state has not yet been realized in monolayer graphene, highlighting new opportunities in the graphene/$\mathrm{PbI}_2$ for exploring spin-orbit-enhanced topological phases in 2D materials.

\section{RESULTS AND DISCUSSION}
Schematic and optical images of a representative graphene/$\rm PbI_2$ device is shown in Fig. \ref{sample}a and b.
The $\rm PbI_2$ flake was prepared by solution epitaxy \cite{sun2020engineering}, with a thickness of around 3 nm (4 layers) and a lateral size of around 10 $\rm \mu m$. Due to its hexagonal symmetry, $\mathrm{PbI}_2$ naturally grows into triangular or hexagonal flakes with well-defined crystallographic edges. Mechanically exfoliated monolayer graphene was patterned into a Hall bar structure and transferred onto the $\mathrm{PbI}_2$ surface using a dry-transfer technique carried out in an argon-filled glovebox. The full heterostructure is encapsulated with a large BN flake. The twist angle between the graphene and the top BN layer is approximately $8^\circ$, while the twist angle between graphene and the underlying $\mathrm{PbI}_2$ is about $31^\circ$. Given the lattice mismatch and relative orientations among the three layers (BN/graphene/$\mathrm{PbI}_2$), a moiré wavelength of approximately 16.9 nm is expected (Supporting information section I). Au/Ti (60 nm/5 nm) are deposited as electronic contacts. The sample is laid on a 285-nm-thick $\rm SiO_2$ layer over doped silicon that serves as a back gate. Additional device geometry, moiré wavelength simulation and characterization are provided in Supplementary information section I. Magnetotransport measurements were performed in a dilution refrigerator at a base temperature of 10 mK using a 9 T superconducting magnet, with standard low-frequency AC lock-in techniques

The carrier density of graphene in the BN/graphene/$\mathrm{PbI}_2$ heterostructure is approximately $1.1 \times 10^{12}\ \mathrm{cm}^{-2}$ at gate voltage Vg = 0 V and the extracted carrier mobility reaches $\sim 14\ 000\ \mathrm{cm^2 V^{-1} s^{-1}}$, indicating a high quality device (Supplementary information section I). Only the graphene directly above the crystalline $\mathrm{PbI}_2$ flake exhibits such high mobility. Residual graphene areas that do not fully cover the $\mathrm{PbI}_2$ exhibit hole doping of $\sim 7.2 \times 10^{12}\ \mathrm{cm^{-2}}$ and no clear CNP is observed within the accessible gate voltage range,  likely due to $\mathrm{PbI}_2$ nanoparticles embedded during the solution-based crystallization process \cite{liu2019few}. 

\subsection{Vanishing Longitudinal Resistance at the CNP}

Fig. \ref{sample}c presents a well-defined Landau fan diagram of the longitudinal resistance $\rm R_{xx}$ from four-probe measurements, with current applied from contact $ a$ to $ b$ and voltage detected between contacts $h$ and $i$, as a function of gate voltage and magnetic field at 10 mK. The CNP is observed at around $\rm V_g =-13.7 \ V$, and the resistance drops to zero between the quantized Landau levels following the expected graphene sequence of Chern number $\nu_h$= ±2, ±6, ±10, ... in quantum Hall regime, which originates from the fourfold spin and pseudo-spin (valley) degeneracy of Landau levels and the non-trivial Berry phase. 

\begin{figure*}[ht]

\includegraphics[clip=true,width=0.8\textwidth]{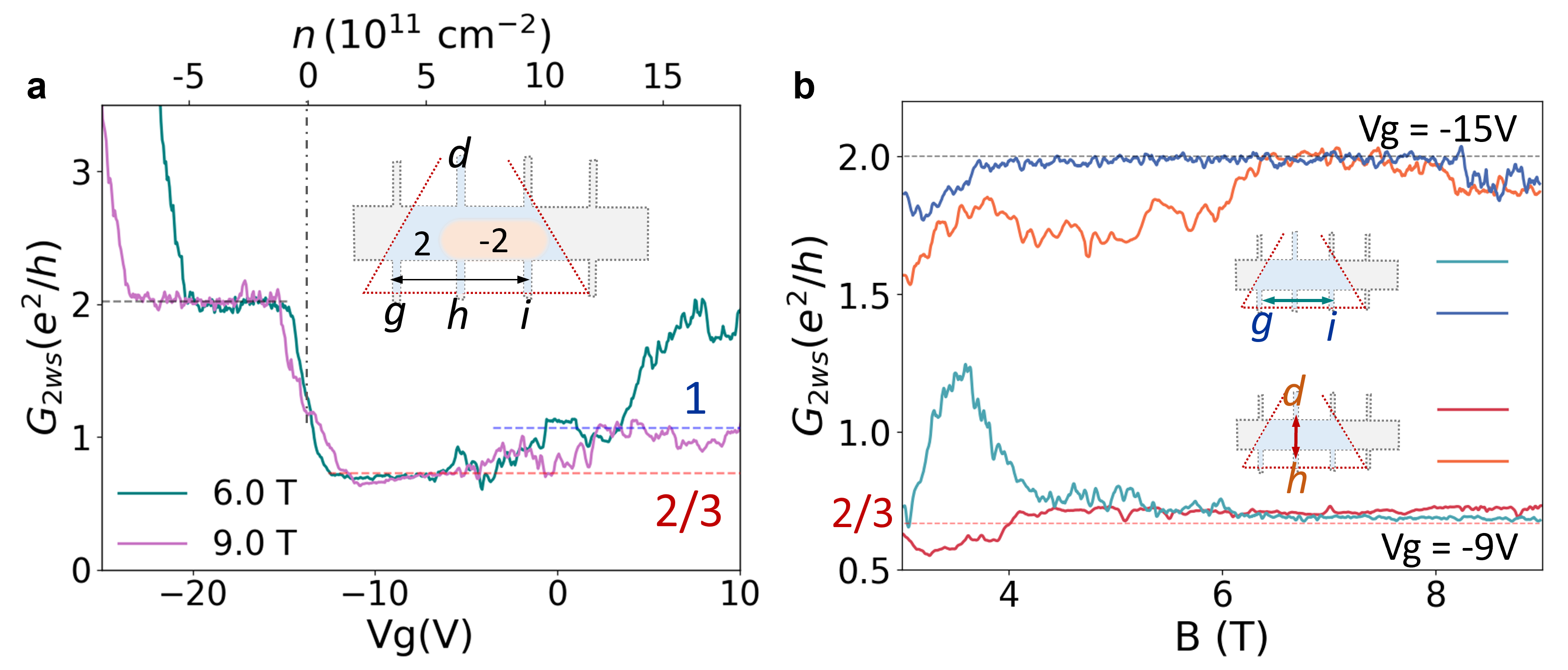} 

\caption{a) Two-probe conductance as a function of gate voltage measured from contacts g–i at magnetic fields of 6 T and 9 T. Well-quantized plateaus at G = $2/3\ e^{2}/h$ and $e^{2}/h$ are observed on the electron-doped side, and $2\ e^{2}/h$ on the hole-doped side. The quantization at $2/3$ and $1$  is consistent with junctions between domains with filling factor $2/-2/2$ and $2/-2$, respectively. The inset displays a schematic of the device geometry and contact configuration, with color-coded regions representing local filling factors. b) Two-probe conductance as a function of magnetic field for two contact configurations, h–d and g–i. A clear $2/3$ conductance plateau appears on the electron side ($\rm V_g = -9\ V$), while a integer plateau at $\nu_h = -2$ is observed on the hole side ($\rm V_g = -15\ V$).   } 
\label{FracPlatplot}
\end{figure*}

Notably, the $\rm R_{xx}$ map in Fig.~\ref{sample}c exhibits anomalies that deviate from conventional expectations for monolayer graphene in the QH regime. Most strikingly, the resistance peak typically associated with the CNP gradually decreases with increasing magnetic field and disappears above $\sim 4.5$ T, giving rise to a broad region of zero longitudinal resistance that extends from the $\nu_h = -2$ to $\nu_h = +2$ plateau across a wide gate voltage range, as shown in Fig.~\ref{sample}d. In addition, a dissipationless region is also observed near the transition between the $\nu_h = 2$ and $\nu_h = 6$ states. The vanishing $\rm R_{xx}$ at CNP and the boundary of $\nu_h$ = 2 and 6 shows linear cutting-edge which intersect zero magnetic field at $n_1$ and $n_2$. We later show how those lines naturely arise from moiré structure.
This unconventional quantum transport also appears in the form of an exceptionally wide $\nu_h = \pm 2$ QH plateau,  which spans a large magnetic field range. For example, at $V_g = -15\ \mathrm{V}$, slightly to the hole-doped side of the CNP, the $\nu_h = -2$ state emerges as early as $B = 1\ \mathrm{T}$ and persists up to the maximum available field of 9 T, forming a remarkably extended plateau. We have ruled out the charge transfer explanation\cite{janssen2011anomalously, kudrynskyi2017giant, kopylov2010charge, tao2020giant}, which fails to account for the persistence of ballistic transport and vanishing $\rm R_{xx}$ across $\nu_h = 0$ at high magnetic fields (Supporting information section II). 

The broad $\rm R_{xx} = 0$ region spanning the CNP, observed between contacts $h$ and $i$, suggests the presence of ballistic transport channels bridging a smooth transition from $\nu_h = +2$ to $\nu_h = -2$.
We suspect that the ballistic channels originate from QH states modulated by a moiré superlattice, in which coexisting topological states are stabilized across different regions of the sample. In such a system, the moiré potential can induce spatially varying minibands, enabling long-range coherent transport that persists even away from CNP \cite{shi2020electronic,bocarsly2025coulomb}. These moiré-induced edge boundary channels offer a possible mechanism for the observed dissipationless behavior \cite{barrier2024one}. 

\subsection{Emergence of a 2/3 Conductance Plateau}

To probe the local transport characteristics of the moir\'e structure, we perform two-probe resistance measurements using electrodes positioned on the graphene/$\mathrm{PbI}_2$ stack. Fig.~\ref{FracPlatplot}a present the conductance between contacts g and i as function of gate voltage at 6 T and 9 T. This shows an asymetric conductance quantization between the electron and hole sides. The conductance remains quantized at $G = 2\ e^2/h$ for the hole density $n_g$ as low as $1 \times 10^{11} \mathrm{cm}^{-2}$ for both $|B|$= 6 and 9 T. At 9 T, this corresponds to a filling factor $\phi_0 n_g / |B| \simeq 0.5 $, for which we no longer expect quantization. The conductance  then  abruptly drops to $G = 2/3\ e^2/h$ near the CNP, forming a well-defined fractional plateau that spans a wide range of carrier densities up to $8 \times 10^{11} \mathrm{cm}^{-2}$ (Fig.~\ref{FracPlatplot}a). As the gate voltage is further increased toward the electron-doped side, an additional integer QH plateau emerges at G = $e^{2}/h$. Figure~\ref{FracPlatplot}b confirms the presence of robust fractional and integer conductance plateaux at $G = 2/3\ e^{2}/h$ and $G = 2\ e^{2}/h$, respectively, as a function of magnetic field. Similar features are observed from both $h$–$d$ and $g$–$i$ contact pairs.

\begin{figure*}[ht]

\includegraphics[clip=true,width=1.0\textwidth]{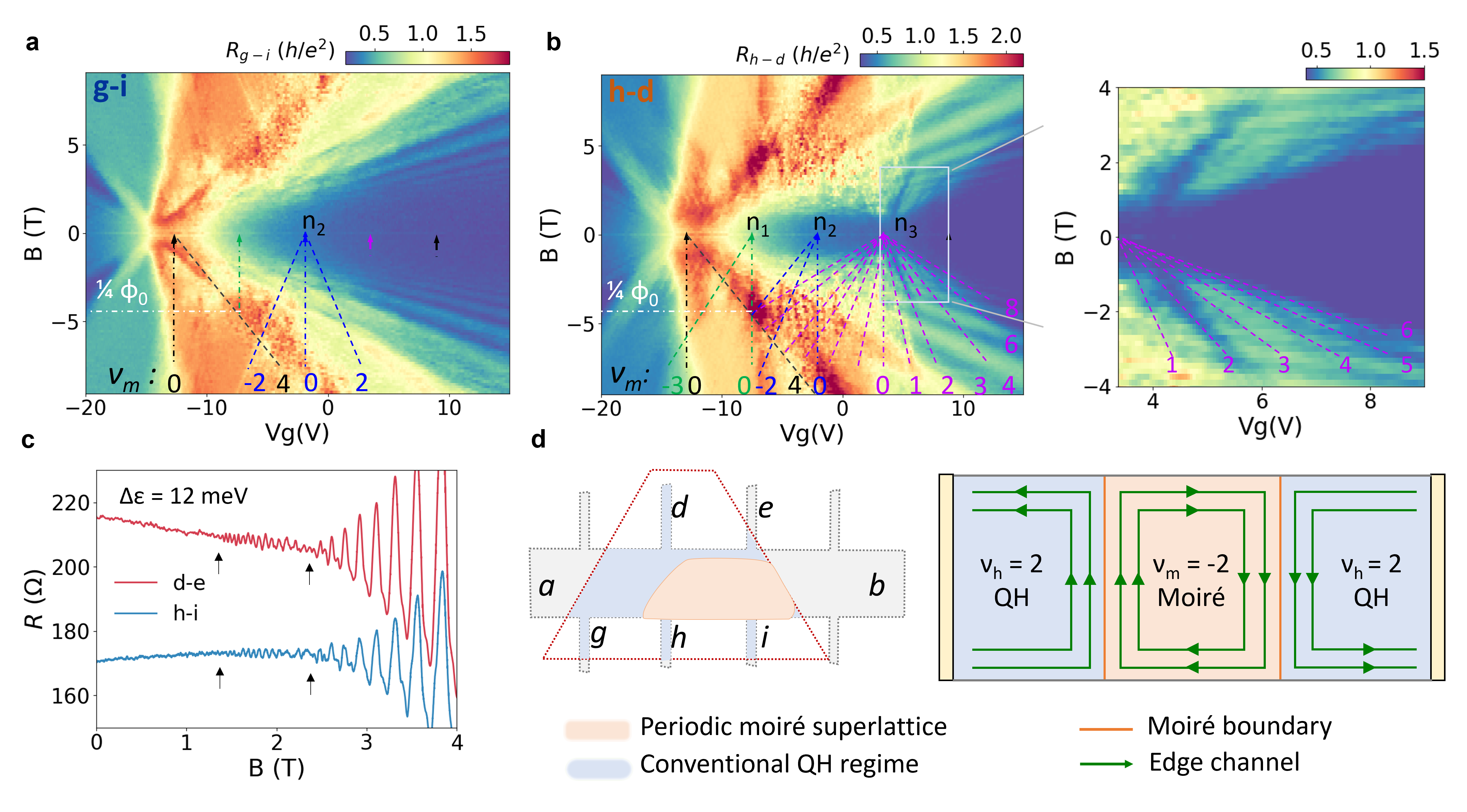} 

\caption{ Two-probe resistance measured as a function of magnetic field and gate voltage for contact pairs (a) $g$–$i$ and (b) $h$–$d$. A clear fractional conductance plateau at $2/3$ emerges on the electron-doped side for $B > 4.5\ \mathrm{T}$, corresponding to a magnetic flux $\phi  \approx \phi_0/4$ (indicated by white dotted lines). The overlaid dashed lines correspond to the experimentally determined Wannier diagram, showing incompressible states that satisfy $n/n_0 = \nu_m \phi/\phi_0 + s_m$. Black dashed lines mark states of ($\rm \nu_m = 0\ and\ 4$, $s_m = 0$), denoting the boundaries of the conventional $\nu_h = 2$ QH plateau in monolayer graphene (degenerate). Blue dashed lines consistent with $\nu_m = \pm 2$, intersect $B = 0$ at $n_2 = 8.62 \times 10^{11}\ \mathrm{cm}^{-2}$ ($s_m = 2$). Purple dashed lines highlight distinct Landau fan structures emerging from $n_3 = 12.93 \times 10^{11}\ \mathrm{cm}^{-2}$ ($s_m = 3$). Green dashed lines indicate the position of $n_1 = 4.31 \times 10^{11}\ \mathrm{cm}^{-2}$ ($s_m = 1$). The numbers label the $\nu_m$ of each linear trajectory. A magnified view of b) (highlighted by a white box) shows $\nu_m = \pm 1, \pm 2, \pm 3, \pm 4$ emerging from $n_3$, indicative of full symmetry breaking.  c) Shubnikov–de Haas oscillations display clear beating patterns, a hallmark of spin–orbit splitting, from which an estimated SOC energy of $\sim 12$ meV is extracted. Arrows mark the oscillation nodes. d) Schematic of the Chern junction naturally formed in the sample. The blue regions represent the conventional QH regime with Chern number of $+2$, while the orange region corresponds to a periodic moiré superlattice domain with a Chern number of $-2$. At the moiré boundary, edge channel mixing leads to current partitioning, which gives rise to the observed fractional quantized conductance of $2/3$. } 
\label{FracPlatmap}
\end{figure*}

A fractional conductance plateau at $2/3\ e^2/h$ has previously been reported in p–n–p junctions formed between $\nu_h = +2$ and $\nu_h = -2$ quantum Hall (QH) domains, where carrier polarity and density are precisely tuned using local electrostatic gates \cite{abanin2007quantized, ozyilmaz2007electronic, williams2007quantum}. In contrast, we propose that the observed $2/3$ conductance plateau in our device originates from naturally formed junctions between moiré domains with distinct Chern numbers. The spontaneous formation of moiré superlattice domains has been reported in twisted bilayer graphene \cite{uri2020mapping,yoo2019atomic}, supporting the existence of twist angle gradients. In our BN/graphene/PbI$_2$ heterostructure, the spatial modulation in the moiré potential creates regions characterized by different Chern numbers, $\nu_h = 2$ and $\nu_m = -2$, leading to the formation of a 2/-2/2 Chern junction that yields the observed $2/3$ conductance quantization. Unlike electrostatically defined p–n junctions, this Chern junction arises intrinsically from the moiré potential landscape, reflecting a spatially modulated electronic topology governed by the underlying moiré physics.

In the two-terminal configuration between contacts h–i, additional conductance plateaus are observed at 10/11 $e^2/h$ and 5/3 $e^2/h$, which can be attributed to possible 2/-10/2 and 2/-10 Chern junctions, respectively. These features disappear at magnetic fields B $>$ 4.5 T, as shown in the Figure S10.  In the same configuration, a pronounced high-resistance peak emerges at carrier density $\rm n_1\ =\ 4.31\ \times \ 10^{11}\ cm^{-2}$, corresponding to the first flavor of the moiré superlattice miniband, at B $>$ 4.5 T. On the hole-doped side of this feature, the conductance becomes quantized at $e^2/h$, indicating that electronic transport is dominated by the moiré Hofstadter miniband. This behavior is consistent with our proposed scenario involving the formation of Chern junctions.

\subsection{Signatures of the Moiré Hofstadter Spectrum}

This moiré-induced modulations is particularly obvious in measurements across the $h$–$d$ contact pair, located entirely in the graphene/$\mathrm{PbI}_2$ region. In this configuration, we observe a series of interpenetrating incompressible states, which obey the Diophantine equation: $\frac{n}{n_0} = \nu_m \frac{\phi}{\phi _0} + s_m$, where $\nu_m$ and $s_m$ are integers, corresponding to the Chern number of filled minibands and the moiré band filling index, respectively \cite{wannier1978result,dean2013hofstadter,wang2015evidence,shen2020correlated}. Here, $n_0$ is the inverse of the moir\'e unit cell area, related to the moir\'e wavelength $\lambda_m$ by $ \lambda_m^2= 2/(\sqrt{3} n_0)$.

In Fig. \ref{FracPlatmap} a, b, the dashed lines represent the experimentally determined Wannier diagram, capturing the evolution of incompressible states as a function of magnetic field and gate voltage \cite{wannier1978result, claro1979magnetic, rothstein2024band}. These states, observed predominantly on the electron-doped side, can be classified based on their zero-field density intercepts: $n_3 = 12.93 \times 10^{11}\ \mathrm{cm}^{-2}$, $n_2 = 8.62 \times 10^{11}\ \mathrm{cm}^{-2}$, $n_1 = 4.31 \times 10^{11}\ \mathrm{cm}^{-2}$. The $n_1$, $n_2$, and $n_3$ form an equally spaced sequence in carrier density. Assuming that $n_1$ corresponds to one electron per moiré unit cell ($s_m = n_1/n_0 = 1$), we extract a moiré wavelength $\lambda_m  \approx 16.37$ nm, which exceeds that expected from individual graphene/BN or graphene/$\mathrm{PbI}_2$ bilayer moiré superlattices. This suggests that the observed periodicity arises from the combined moiré potential formed in the BN/graphene/$\mathrm{PbI}_2$ stack with the specific twist angles in our device. Quantized Hall plateaus at $\nu_h = \pm 2,\ \pm 4,\ \pm 6,\ldots$ are observed emanating from the CNP, though no signatures of spin or valley degeneracy lifting are apparent in this region. 

Different from commonly reported four-flavor structure in graphene moiré superlattices, our measurements reveal three well-defined moiré flavors, with the feature at $n_3$ being prominent (Fig.~\ref{FracPlatmap}b). The signature of $n_2$ is shown in the geometry of $g-i$ (Fig.~\ref{FracPlatmap}a), where a linear trajectory associated with $\nu_m = \pm 2$ intersects the $ 2/3$ fractional plateau and extrapolates to a zero-field intercept at $n_2$. This suggests a relation between the underlying moiré miniband structure and the stabilization of the fractional state. The $n_1$ feature is characterized by resistive $\nu_m = 0$ states, which act as the boundary of $G = 2/3\ e^2/h$ (Fig.~\ref{FracPlatmap}b).

A distinct set of Landau levels emerges from $n_3$ under intermediate magnetic field, where we see incompressible QH states with $\nu_m = \pm 1, \pm 2, \pm 3, \pm 4$ (Fig.~\ref{FracPlatmap}b), indicating robust miniband formation. While these states are weaker in amplitude than those originating from the CNP, their presence indicates full symmetry breaking of the moiré minibands. However, only a weak secondary Dirac peak is detectable at $n_3$ in the zero-field resistance (Fig.~S9), likely reflecting the spatial inhomogeneity of the moiré potential. This is consistent with the mechanical flexibility of $\mathrm{PbI}_2$ nanosheets \cite{ran2019mechanical}, which may produce local variations in the superlattice potential that are inconspicuous at zero magnetic field but become amplified in the QH regime, where edge channel dominates transport. 

We suspect that unusual sequence of $\nu_m$ and $s_m$ identified in our Wannier diagram is influenced by the strong SOC induced by the $\mathrm{PbI}_2$ layer. Analysis of the Shubnikov–de Haas beating patterns at higher carrier densities yields an estimated SOC strength of approximately 12 meV (Fig.~\ref{FracPlatmap}c and Fig. ~S13), supporting the presence of proximity-induced spin–orbit interaction \cite{wang2016origin,nitta1997gate}. The strong SOC plays a key role in shaping the miniband structure and the resulting transport characteristics of the system.

The moiré superlattice appears to play a central role in the emergence of the $2/3$ fractional conductance. In particular, the trajectories derived from the moiré band structure delineate the boundaries of the $2/3$ region. The onset of the $ 2/3$ fractional plateau at magnetic fields above $B > 4.5\ \mathrm{T}$ corresponds to a flux density of approximately $\phi_0/4$ per moiré unit cell, by taking the measured $\lambda_m = 16.37$ nm. The fractional plateau is bounded on one side by the lines ($\rm \nu_m = 0\ and\ 4$, $s_m = 0$), which mark the termination of the degenerate $\nu_h = 2$ state on the electron side in monolayer graphene. On the other side, it is enclosed by the states corresponding to ($\nu_m = -2,\ s_m = 2$) and ($\nu_m = 0,\ s_m = 1$). These features indicate that the $2/3$ state arises from a junction with Chern numbers of 2/-2/2, where $\nu_h = 2$ corresponds to the conventional integer QH regime with negligible influence from the moiré potential, and $\nu_m = -2$ originates from periodic moiré superlattice, as illustrated in Fig.~\ref{FracPlatmap}d. This moiré structure connects the contact pair of $h$–$i$, giving rise to the ballistic transport channel across the CNP. While for both contact pairs of $d $–$h $ and $g $–$i $, the formation of 2/-2/2 Chern junction results in the emergence of the 2/3 fractional conductance. 

While moiré-related features are most prominent in two-probe measurements, signatures are also evident in four-probe resistance (Fig. \ref{sample}c and Fig.~S10). There, incompressible states associated with $n_2$ at $\nu_m = 1, 4, 5, 8, 10$, and with $n_1$ at $\nu_m = 3, 4$  show good correspondence with the vanishing longitudinal resistance in $\rm R_{xx}$. The associated vanishing $\rm R_{xx}$ across the CNP strongly indicates that the moiré superlattice facilitates the formation of ballistic edge-like channels, which mediate dissipationless transport across QH domain boundaries.

The 2/-2/2 Chern junction is likely originated from twist-angle variations and strain-induced dislocations, enabled by the mechanical flexibility of the $\mathrm{PbI}_2$ nanosheets. These structural inhomogeneities can introduce spatial gradients near the boundary of the moiré superlattice. Within the quantum Hall regime, graded moiré domain walls can act as beam splitters, giving rise to resistance fluctuations whose amplitude reflects the transmission probabilities of edge states across these interfaces. A hallmark of this mechanism is the correlated appearance of resistance peaks on opposite edges of the device, suggesting the presence of coherent transport that bridges the two sides.  To explore this possibility, we perform four-probe resistance measurements across opposite edges of the Hall bar, focusing on correlations in resistance fluctuations. 

\begin{figure*}[ht]

\includegraphics[clip=true,width=1.0\textwidth]{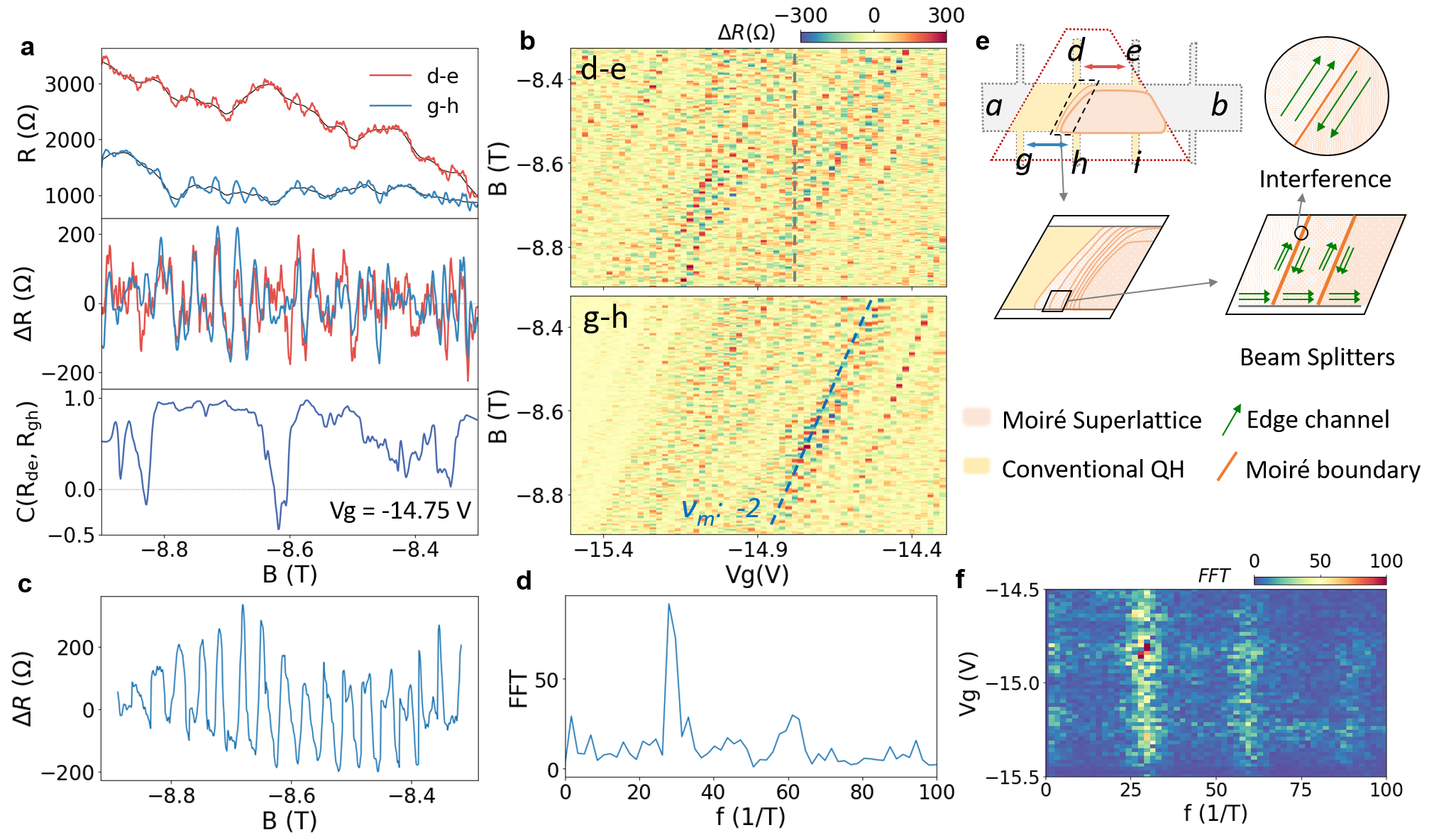} 

\caption{Correlated resistance fluctuations in BN/graphene/$\mathrm{PbI}_2$ heterostructure.  a) Top: Magnetoresistance measured from spatially separated contact pairs ($d $–$e $ and $ g$–$h $) at fixed gate voltage $V_g = -14.75\ \mathrm{V}$ show pronounced resistance fluctuations in the quatized region. Middle: The resistance fluctuations $\Delta R$ isolated by subtracting low-frequency background. Bottom: Covariance of the $\Delta R$ between contact pairs of $d $–$e $ and $g $–$h $ reveals strong correlation in selected magnetic field windows, indicating non-trivial coherence effects.  b) $\Delta R$ as a function of gate voltage and magnetic field for the $d $–$e $ (top) and $ g$–$h $ (bottom) contact pairs. The fluctuation peaks follow linear trajectories with the $\nu_m = -2$. The grey dashed line marks the slice at $V_g = -14.75\ \mathrm{V}$ shown in (a), and the blue dashed line follows one trajectory of $\nu_m = -2$. c) Extracted $\Delta R$ along the blue line in (b) , plotted as a function of magnetic field, exhibits periodic oscillations characteristic of quantum interference. d)  Fourier transform of the oscillations in (c), revealing a dominant frequency at $f = 28.18\ \mathrm{T}^{-1}$, corresponding to an interference area of $\sim$0.116 $\mu\mathrm{m}^2$.  e) The schematic illustrates that  graded moiré domain walls form at the boundary of moiré superlattice, acting as electronic beam splitters and leading to the synchronised resistance fluctuations from contact pairs of  $d $–$e $ and $ g$–$h $. A portion of the edge channel following the moiré boundary interferes with counter-propagating edge states, resulting in a electronic interference pattern. f)  Fourier amplitude as a function of gate voltage, exhibits enhanced oscillatory features near $\rm V_g \approx -14.85\ V\ and\ -15.25\ V$, coinciding with pronounced resistance fluctuations, which indicates the origin in electronic interference from coherent transport modulated by moiré domain walls.} 
\label{fluctuation}
\end{figure*}

\subsection{Correlated Resistance Fluctuations and Quantum Interference}
Within the $\nu_h = -2$ QH regime, we observe spatially correlated resistance fluctuations between the $d$–$e$ and $g$–$h$ contact pairs. Fig.~\ref{fluctuation}a illustrates the extraction of these fluctuations from the magnetoresistance. In the top panel, the raw four-probe magnetoresistance for both contact pairs are shown. The middle panel shows the fluctuation components $\Delta R$, obtained by subtracting the low-frequency background from the original signals. These resistance fluctuations yield a covariance close to 1 within the magnetic field intervals of $-8.81\ \mathrm{T}$ to $-8.66\ \mathrm{T}$ and $-8.58\ \mathrm{T}$ to $-8.49\ \mathrm{T}$ at gate voltage of $V_g = 14.75\ \mathrm{V}$ (bottom panel), indicating a strong correlation between contact pairs, $d$–$e$ and $g$–$h$. 
The presence of clear inter-edge correlations between macroscopically separated electrodes distinguishes these fluctuations from conventional universal conductance fluctuations \cite{poumirol2010electron, martin2008observation, yang2016puddle, qiu2022mesoscopic}, instead indicating a fundamentally different, moiré-modulated coherent transport mechanism.
The fluctuation amplitude is around $\rm R_f \approx 150\ \Omega$. Assuming these fluctuations result from partial backscattering across moiré gradient boundaries, we estimate the transmission probability of QH edge states across these boundaries as $\rm  T \approx 1 - R_f / R_Q \sim 99.4\%$, where $\rm R_Q = h/e^2$ is the resistance quantum.

Fig.~\ref{fluctuation}b shows the $\Delta R$ as a function of gate voltage and magnetic field. The correlated resistance fluctuations follow linear trajectories consistent with $\nu_m = -2$, confirming that they originate from quantized edge modes. This map corresponds to a $\nu_h = -2$ QH state near the CNP, indicating that both sides of the moiré domain boundaries are quantized to the same Chern number, $\nu_m =\nu_h = -2$. 

The transport along the moiré boundaries  is expected to be accompanied by interference-induced oscillation in $\rm R_{xx}$. Indeed, superimposed on the fluctuation background in Fig.~\ref{fluctuation}b, we observe oscillatory features, similar to those observed in QH electronic interferometers \cite{jo2021quantum, wei2017mach, deprez2021tunable, jo2022scaling, werkmeister2024strongly,chakraborti2025electron}. In the QH regime, such interference arises from edge states forming coherent loops when partially transmitted and reflected by quantum point contacts or junctions, such as p–n interfaces or moiré domain boundaries. 

To further investigate the interference signatures, we extract $\Delta R$ along a $\nu_m = -2$ trajectory, marked by the blue dashed line in Fig.~\ref{fluctuation}b, and track its evolution with magnetic field. As shown in Fig.~\ref{fluctuation}c, the $\Delta R$ exhibits clear periodic oscillations. A Fourier transform of this oscillation reveals a dominant frequency peak at $f = 28.18\ \mathrm{T}^{-1}$, along with a visible second harmonic (Fig.~\ref{fluctuation}d). This corresponds to an effective interference area of approximately 0.116 $\mu m^2$ . Considering the device width of $\sim 5\ \mu$m, the inferred separation between interfering edge paths is $\sim 23$ nm, which is in the same order as the estimated moiré wavelength in our sample ($\lambda_m \approx 16.3$ nm). These observations confirm that the quantum oscillations originate from coherent transport modulated by moiré domain walls.

The drawing in Fig.~\ref{fluctuation}e illustrates a spatially graded moiré domain that emerges between the periodic moiré superlattice and the conventional QH regime. These graded moiré domain walls split electronic edge channels, leading to the correlated resistance fluctuations observed across contact pairs  $d $–$e $ and $g$–$h$. The edge modes guided along the moiré boundary, interfere with counter-propagating edge modes, giving rise to a distinct electronic interference pattern superimposed on the magnetoresistance fluctuations. 

In Fig.~\ref{fluctuation}f, the Fourier amplitude as a function of gate voltage reveals enhanced oscillatory features near $\rm V_g \approx -14.85\ V\ and\ -15.25\ V$, where the resistance fluctuations are more pronounced, reinforcing the connection between magnetoresistance fluctuations and electronic interference. By extrapolating the $\nu_m$= -2 linear trajectories of the fluctuation peaks to zero magnetic field, we infer the corresponding local moiré periodicities, which provide insights into the characteristics of specific moiré domains. The extracted fluctuation lines correspond to subtle variations in the moiré wavelength, with $\lambda_m$ ranging from 17.39 nm to 19.97 nm (assuming the intercept density as $n_0$), indicating the presence of a spatial gradient in moiré periodicity across the device. These slight variation in $\lambda_m$ are likely induced by strain or local stress within the mechanically flexible $\mathrm{PbI}_2$ layer, which can continuously deform and modulate the moiré pattern at nanoscale.

\section{CONCLUSION}

In conclusion, we report QH effect in a moir\'e heterostructure BN/graphene/$\rm PbI_2$. Our measurements reveal a continuous dissipationless region with $\rm R_{xx} = 0$ that persists across the CNP ($\nu_h = 0$), bridging the transition between $\nu_h = -2$ and $\nu_h = 2$ QH states, which indicates the presence of a strong moir\'e potential. Moiré features are clearly manifested in two-probe resistance measurements, which reveal an extended region exhibiting fractional quantization at $ 2/3$ and an unconventional pattern of moiré miniband flavors. We attribute the unusual flavor sequence to a strong proximity-induced spin–orbit interaction, supported by the observation of beating patterns in Shubnikov–de Haas oscillations. We suggest that persistence of the ballistic channel at quantum hall phase boundaries is a combined effect from the Kane-Mele type of SOC induced by the $\mathrm{PbI}_2$ and the moir\'e super-potential giving rise to a high magnetic field topological insulator phase in a monolayer graphene heterostructure.

The fractional $2/3$ conductance plateau arises from a junction between integer QH states and moiré-induced topological miniband with distinct Chern numbers $\pm 2$. The boundaries of the 2/3 plateau are captured by linear trajectories characteristic of Hofstadter physics in moiré systems. The mechanical flexibility of the $\mathrm{PbI}_2$ nanosheets leads to local strain and gradients in the moiré superlattice period, giving rise to domain walls that act as quantum beam splitters. The current scattered at the graded moiré domain walls generates coherent resistance fluctuations, correlated across macroscopic distances, and follow line with $\nu_m = -2$ in the magnetic field–gate voltage plane, which exhibit quantum interferometric behavior in QH regime.
These findings highlight the potential of graphene/$\mathrm{PbI}_2$ heterostructures as a versatile platform for engineering correlated and topological quantum phases, particularly with moiré superlattices.

\vspace*{5mm}
{\bf Acknowledgements:} We acknowledge S. Guéron, C. Quay H. L., R. Deblock,  P. Roulleau, C. Morice for helpful discussions on the manuscript. This project was supported by funding from ANR-20-CE92-0041 (MARS) and IDF DIM QuanTiP, and the European Research Council (ERC) under the European Union’s Horizon 2020 research and innovation program (grant Ballistop agreement no. 833350). The data in this article are available from the authors upon reasonable request.

\bibliography{gpi_cp.bib}

\clearpage 
\newpage
\onecolumngrid

\section{Sample information }
\vspace*{5mm}
\noindent\textbf{\bf Device Fabrication}

The hBN/graphene/$\rm PbI_2$ heterostructures were fabricated using a polymer-assisted dry-transfer technique designed for precise rotational alignment between layers. Monolayer graphene and thin hBN flakes were first exfoliated onto a  $\rm SiO_2$/Si substrate. A top hBN layer was then picked up and aligned to graphene with a twist angle of $8^\circ$, using a PC/PDMS stamp under an optical microscope equipped with a rotational micro-manipulator. The hBN/graphene stack was subsequently patterned into a Hall bar geometry by reactive ion etching, as shown in Fig. S1b. This Hall bar stack was then picked up and transferred onto a $\rm PbI_2$ crystal with a twist angle of $31^\circ$. The twist angles were determined by aligning the straight crystal edges—typically corresponding to crystallographic axes—yielding an angular precision better than 0.001°. These chosen angles produce a large moiré wavelength ($\sim 16.9$ nm), as predicted by reciprocal-space simulations (Fig. S2), allowing clear observation of moiré-induced effects while minimizing lattice relaxation commonly seen in small-angle graphene/hBN systems.

All transfers were carried out in an argon-filled glovebox to prevent degradation of $\rm PbI_2$. The completed heterostructure was encapsulated by an additional thick hBN layer for environmental stability. Electrical contacts were formed by depositing Ti/Au (5 nm/60 nm) electrodes. The device rests on a 285 nm $\rm SiO_2$ layer atop doped silicon, which serves as the global back gate. The measured transport data confirm a moiré wavelength of approximately 16.4 nm, consistent with the designed geometry and validating the accuracy of the alignment process.

\begin{figure*}[ht]

\includegraphics[clip=true,width=0.9\textwidth]{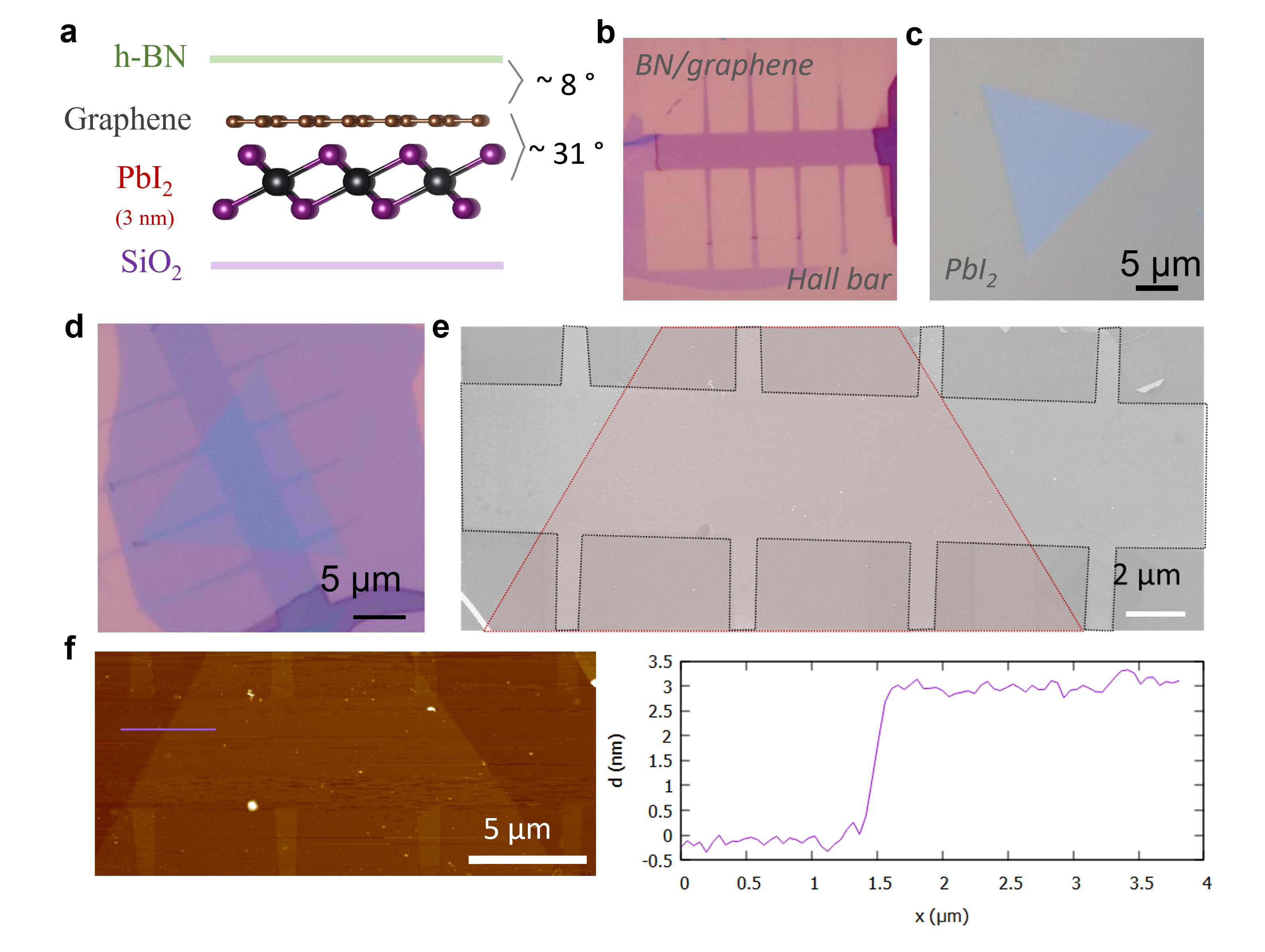} 

\caption{ a) Schematic of the device structure, where monolayer graphene is sandwiched between hBN and $\mathrm{PbI}_2$. The twist angle between graphene and hBN is approximately $8^\circ$, and between graphene and $\mathrm{PbI}_2$ is about $31^\circ$. b) Optical image of the fabricated hBN/graphene Hall bar. A thin hBN flake ($<$3 nm) is dry-transferred onto graphene with an $\sim 8^\circ$ rotational alignment guided by graphene edge orientation. The Hall bar (width $\sim 5\ \mu$m) is defined by reactive ion etching. Pre-etch hBN capping preserves graphene cleanliness and prevents wrinkling during fabrication. c) Optical image of a pristine, triangular $\mathrm{PbI}_2$ single crystal grown via solution epitaxy. d) Optical image of the final hBN/graphene/$\mathrm{PbI}_2$ stack, encapsulated by a 30 nm top hBN flake. e) SEM image of the completed device, showing the full heterostructure layout. f) AFM height profile of the device after transport measurement and a cross-sectional line trace showing the measured thickness of the $\mathrm{PbI}_2$ layer  }
\label{sampleimage}
\end{figure*}

\begin{figure*}[ht]

\includegraphics[clip=true,width=0.8\textwidth]{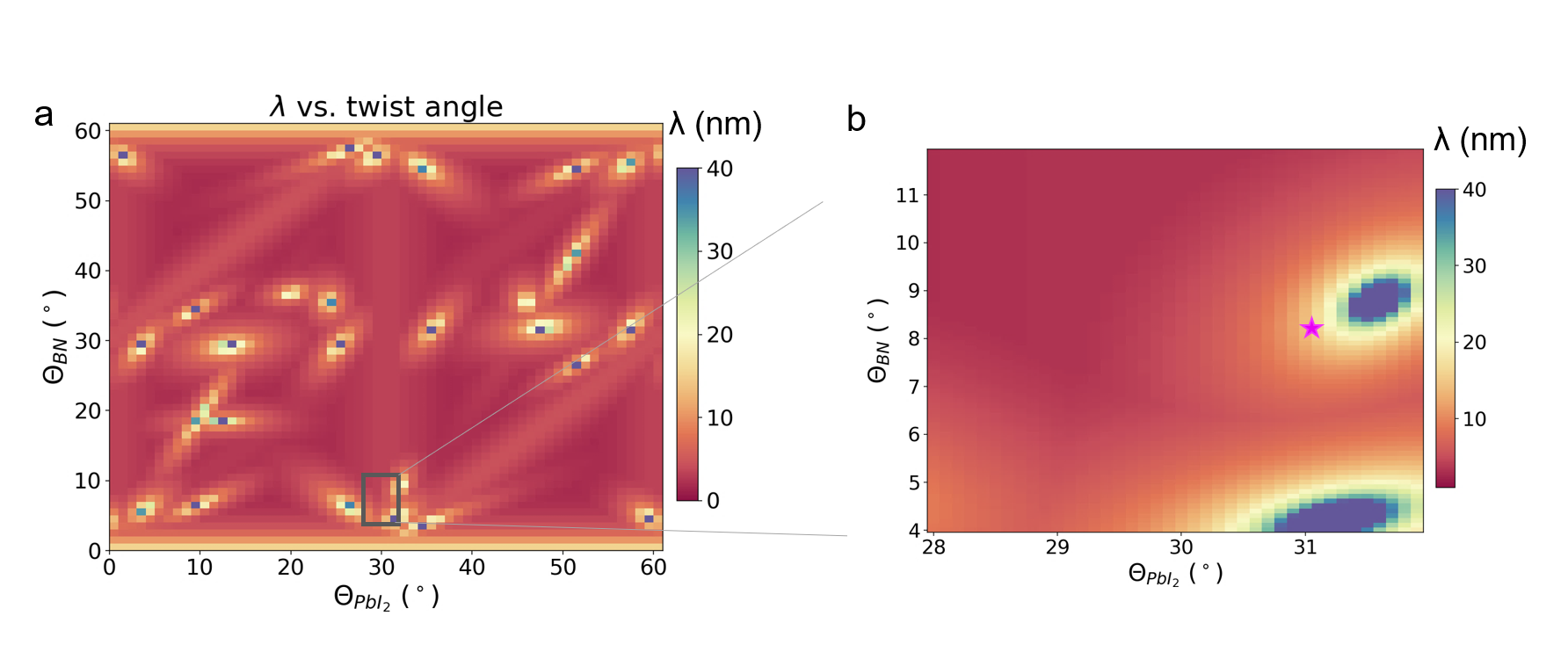} 

\caption{ a) Simulated moiré wavelength as a function of twist angles between hBN and graphene ($\theta_{\mathrm{BN}}$), and $\mathrm{PbI}2$ and graphene ($\theta{\mathrm{PbI_2}}$), calculated by determining the moiré vector in reciprocal space. The map shows that certain angle combinations yield large moiré periods.
b) Zoomed-in region of the map near the experimentally relevant twist angles, indicating that $\theta_{\mathrm{BN}} \approx 8^\circ$ and $\theta_{\mathrm{PbI_2}} \approx 31^\circ$ (marked by the pink star) correspond to a moiré wavelength of approximately 16.9 nm.}
\label{simulation}
\end{figure*}

\begin{figure*}[ht]

\includegraphics[clip=true,width=0.5\textwidth]{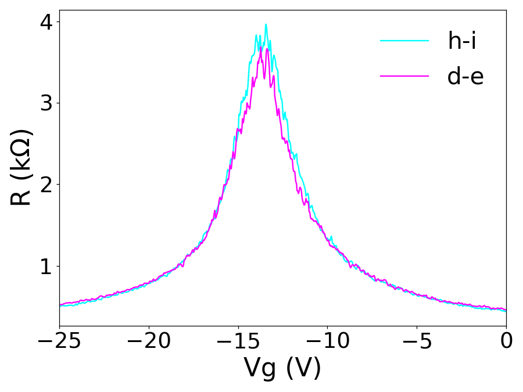} 

\caption{Four-probe resistance measurements at $B = 0\ \mathrm{T}$ and $T = 10\ \mathrm{mK}$ from contact pairs h–i and d–e, with current sourced from contact a to b, shows the charge neutrality point identified near $V_g = -13.7\ \mathrm{V}$.} 
\label{transportcurve}
\end{figure*}

\begin{figure*}[ht]

\includegraphics[clip=true,width=1.0\textwidth]{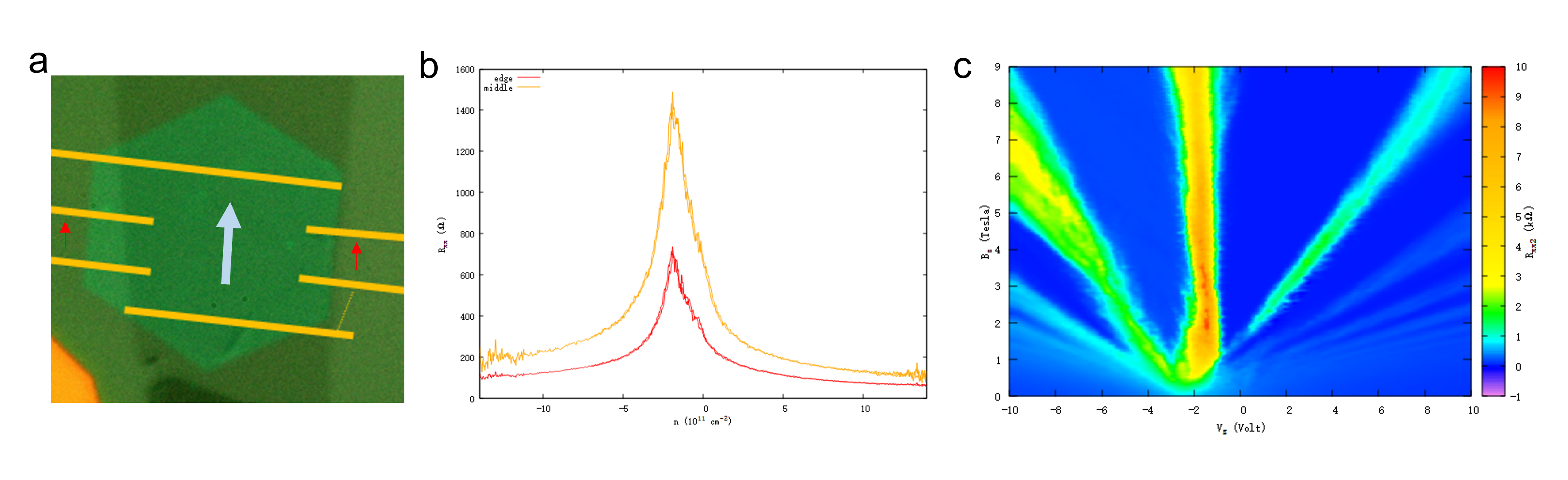} 

\caption{A control device of hBN/Graphene/$\mathrm{PbI}2$ with twist angles $\theta{\mathrm{BN}} \approx 41^\circ$ and $\theta_{\mathrm{PbI}_2} \approx 8^\circ$, corresponding to a moiré superlattice wavelength of $\sim$3.6 nm. Low-field transport measurements reveal a carrier density of $\sim 1.5 \times 10^{11}\ \mathrm{cm}^{-2}$ and mobility of $\sim 40{,}000\ \mathrm{cm^2 V^{-1} s^{-1}}$. c) A well-defined Landau fan diagram and a high-resistance Dirac point are observed. Notably, a relatively lower resistance along the phase boundary on the electron-doped side, emanating from $V_g \sim 0\ \mathrm{V}$, may signal interface-induced spin–orbit coupling in graphene \cite{cysne2018quantum}.  }
\label{Controlsample}
\end{figure*}

\begin{figure*}[ht]

\includegraphics[clip=true,width=0.8\textwidth]{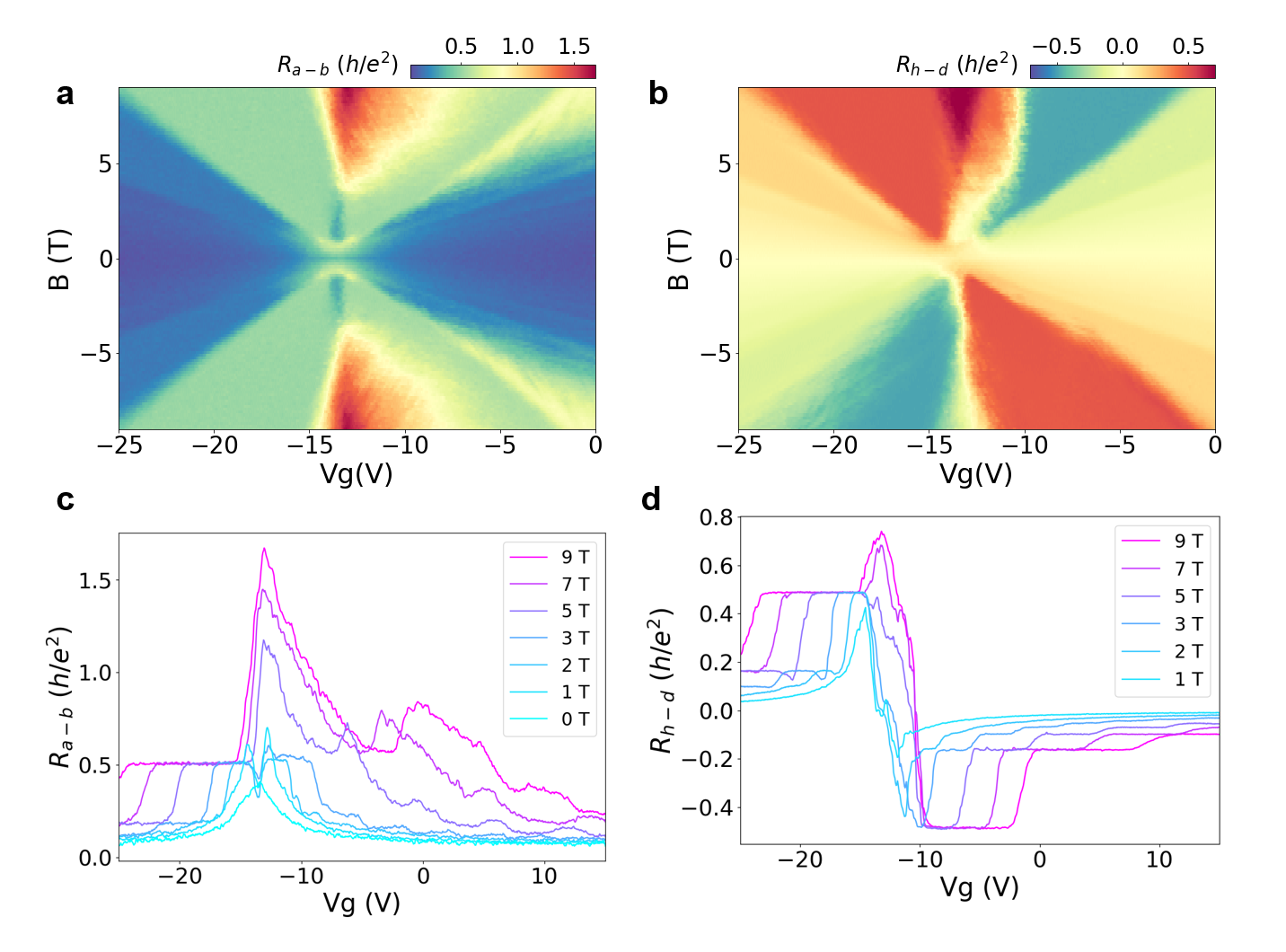} 

\caption{a) Two-probe resistance measured from contact pair 'a–b' exhibits well-quantized Hall plateaus at filling factors $\nu = -2, -6, -10,\ldots$ on the hole-doped side, while a pronounced high-resistance feature appears on the electron-doped side. b) Hall resistance as a function of gate voltage and magnetic field reveals quantized plateaus at $\nu = \pm 2, \pm 6, \pm 10,\ldots$. c, d) Line cuts of magneto-resistance from panels (a) and (b) at various magnetic fields. The zero-crossing point of the Hall resistance gradually shifts toward the electron-doped side as the field increases from 3 T to 6 T. }
\label{R2wsRxy}
\end{figure*}

\begin{figure*}[ht]

\includegraphics[clip=true,width=1.0\textwidth]{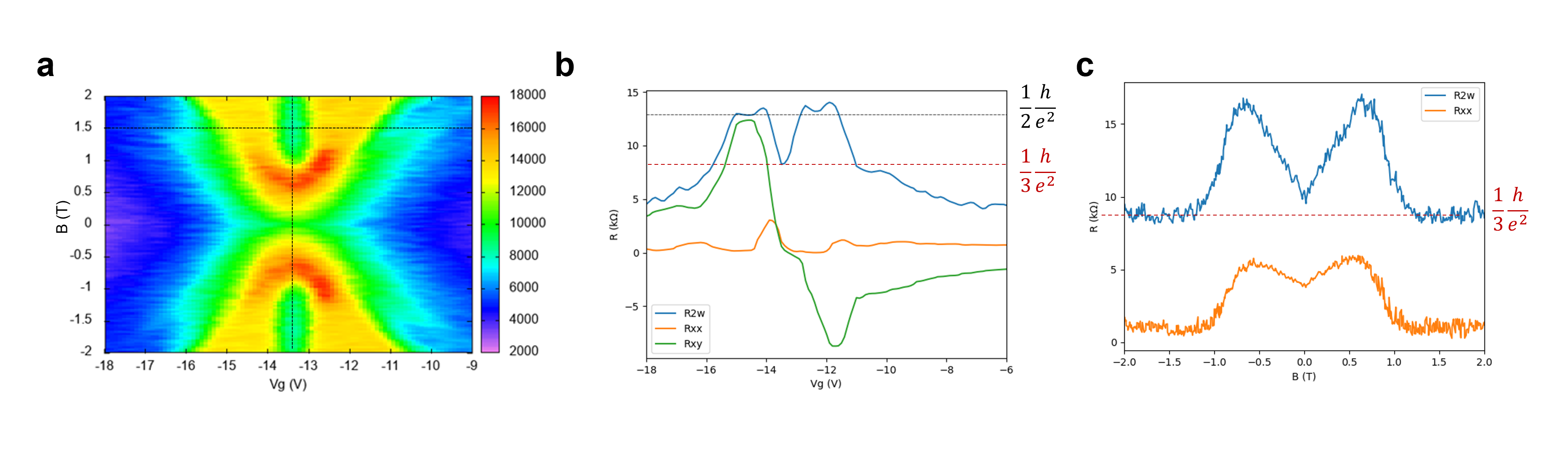} 

\caption{a) At low magnetic fields near the Dirac point, a distinctive "arc"-shaped high-resistance region and a "drop"-shaped low-resistance feature are observed. b) Line cut of resistance versus gate voltage at $B = 1.5\ \mathrm{T}$.
c) Resistance as a function of magnetic field at $V_g = -13.5\ \mathrm{V}$ (close to the Dirac point), showing that the "drop"-shaped feature corresponds to a resistance near $R \sim \frac{1}{3}\ h/e^2$ and occurs where both longitudinal and Hall resistances approach zero. This behavior is likely associated with a parallel-oriented $2/-2$ quantum Hall junction formed near the Dirac point. }
\label{arc}
\end{figure*}

\section{Quantum Hall plateaus }
{\bf Giant quantum Hall plateaus }
Giant quantum Hall plateaus have previously been reported in graphene heterostructures coupled with substrates such as SiC, InSe, and $\mathrm{LaAlO}_3/\mathrm{SrTiO}_3$, where the effect was attributed to charge transfer processes within the heterostructure interface \cite{janssen2011anomalously, kudrynskyi2017giant, kopylov2010charge, tao2020giant}.  In this model, the sample carrier density is reduced on both electron/hole sides, compared to values predicted by the geometrical capacitance, due to the vanishing graphene compressibility at the Dirac point. Under magnetic field, Landau level formation suppresses the compressibility anomaly, enabling the carrier density to increase and approach the value expected from the geometrical capacitance, thereby producing a broadened quantum Hall plateau. According to this model, the maximum magnetic field for observing the $\nu = \pm 2$ quantum Hall states should be constrained by the relation $|B| \leq \phi_0 / (2 n_g)$, where $n_g$ is the carrier density determined from the gate/sample capacitance.  However, it is not consistent with our observation. In addition, we observe no significant deviation in the carrier density versus gate voltage relation, as independently verified from low-field Hall resistance measurements. These two reasons indicate that charge transfer cannot explain the wide quantum Hall effect plateaus in our experiment. The detail analysis is as following.

The physical model to explain the giant $\nu$=2 hall plateau in graphene/SiC considers the quantum capacitance-dominated charge transfer between the two layers. Here we attempt to adopt it into our $\rm Graphene/PbI_2$ heterostructure. This model was proposed by Janssen and Falko \cite{janssen2011anomalously,kopylov2010charge}, and has been applied in the $\rm Graphene/LaAlO_3/SrTiO_3$ system \cite{tao2020giant} and discussed in disordered graphene system \cite{poumirol2010electron,yang2016puddle}.  

The charge balance equation: 
\begin{align}
  {\gamma}[A-\frac{e^2}{C} (n_s+n_g) - \epsilon _F] = (n_s+n_g) 
  \label{eq:cb}
\end{align}  
The left-hand side of this equation accounts for the depletion of the surface donor states, where A is the difference between the work function of undoped graphene and the work function of electrons in the surface donors in SiC, $\epsilon _F$ is the Fermi energy of electrons in graphene, and $\gamma$ is the density of donor states in the dead layer. $n_s$ of this charge density is transferred to graphene, and
$n_g$ is controlled by the gate voltage. The quantum capacitance is believed to be strong as the distance between the donor layer and graphene is short (d= 0.3-0.4 nm), which causes the pinning of the electron filling factor instead of oscillation of electron density. The quantum capacitance takes the form of $C_q = e^2 \gamma _e$, while somehow the form used in this article is the same as classical capacitance $C_c = 1/(4 \pi d)$.

The game played here is the crossover between quantized Landau level expressed by $n_s = \nu B/ \phi_0 $ and Fermi energy - $\epsilon_F = V_F \sqrt {2 e \hbar B |N|}$ for high field and $\epsilon_F = \hbar V_F \sqrt {\pi n_s}$ for 0 field. $v = 4 N +2$. 

\begin{figure}[h]

\includegraphics[clip=true,width=0.8\textwidth]{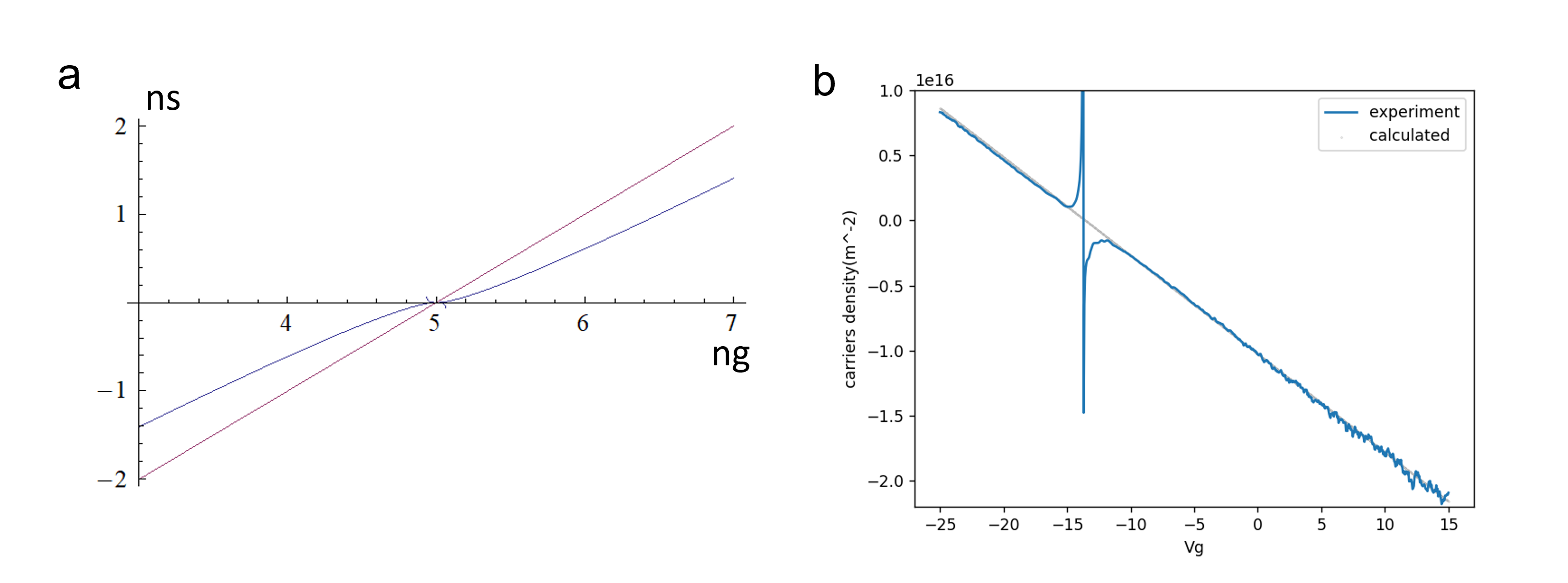} 

\caption{ a) Theoretical relation between the surface carrier density $n_s$ and gate-induced charge density $n_g$ at $B = 0$. The blue curve corresponds to the conventional Dirac dispersion model with Fermi energy $\epsilon_F = \hbar v_F \sqrt{\pi n_s}$. The red curve includes charge trapping effects, modeled by $n_s = \frac{A\gamma}{1 + e^2 \gamma / C} - n_g$, where $\gamma$ is the density of trapped states and $C$ is the gate capacitance per unit area.
b) Experimentally extracted carrier density versus gate voltage from Hall resistance measurements at low magnetic field.} 
\label{nsngB0}
\end{figure}

At low field, we take $\epsilon_F = \hbar V_F \sqrt {\pi n_s}$, the Eq. \ref{eq:cb} can be reformed as $A - \hbar V_F \sqrt {\pi n_s}= (n_s+n_g) (\frac{1}{\gamma} + \frac{e^2}{C}) $.  Comparing with the $n_s$ vs. ${n_g}$ when $\epsilon_F$ = 0,  $n_s ( {n_g})$ shows weaker dependence on the gate voltage when $n_s$ is small, as shown in Figure \ref{nsngB0}a. The deviation increases with the $\sqrt {n_g}$ ($\Delta n_s = \frac{\gamma}{1+e^2/C \gamma}\times \frac{v_F^2 \gamma - \sqrt{v_F^2(4 ng (1+e^2/C \gamma)^2 + (v_F-4A (1+e^2/C \gamma))}}{2(1+e^2/C \gamma)}$).  However, in the experiment, the carrier density extracted from hall resistance changes linearly with the gate voltage and can be reproduced by taking the 285 nm thick $\rm SiO_2$ as the insulator corresponding to the Capacitance of around  $12.3 \ nF/cm^2$. If we take the theoretical deviation $\Delta n_s $ to zero (mimic the experimental results), one gets  $n_g = \frac{A \gamma}{(1+e^2/C \gamma)}$ (or $ \frac{\gamma}{(1+e^2/C \gamma)}=0$, which seems not very reasonable, as it will cause the decay of Eq \ref{eq:cb} to A = $\epsilon_F$ and no gate tunability). At $n_g = 0$, experimentally we get the electron doping level of graphene around $n_s(0) = 1.1 \times 10^{12} cm^{-2}$, which may give the $n_s = \frac {A \gamma}{1+e_2 \gamma / C} = n_s(0)$

\begin{figure}[ht]

\includegraphics[clip=true,width=0.8\textwidth]{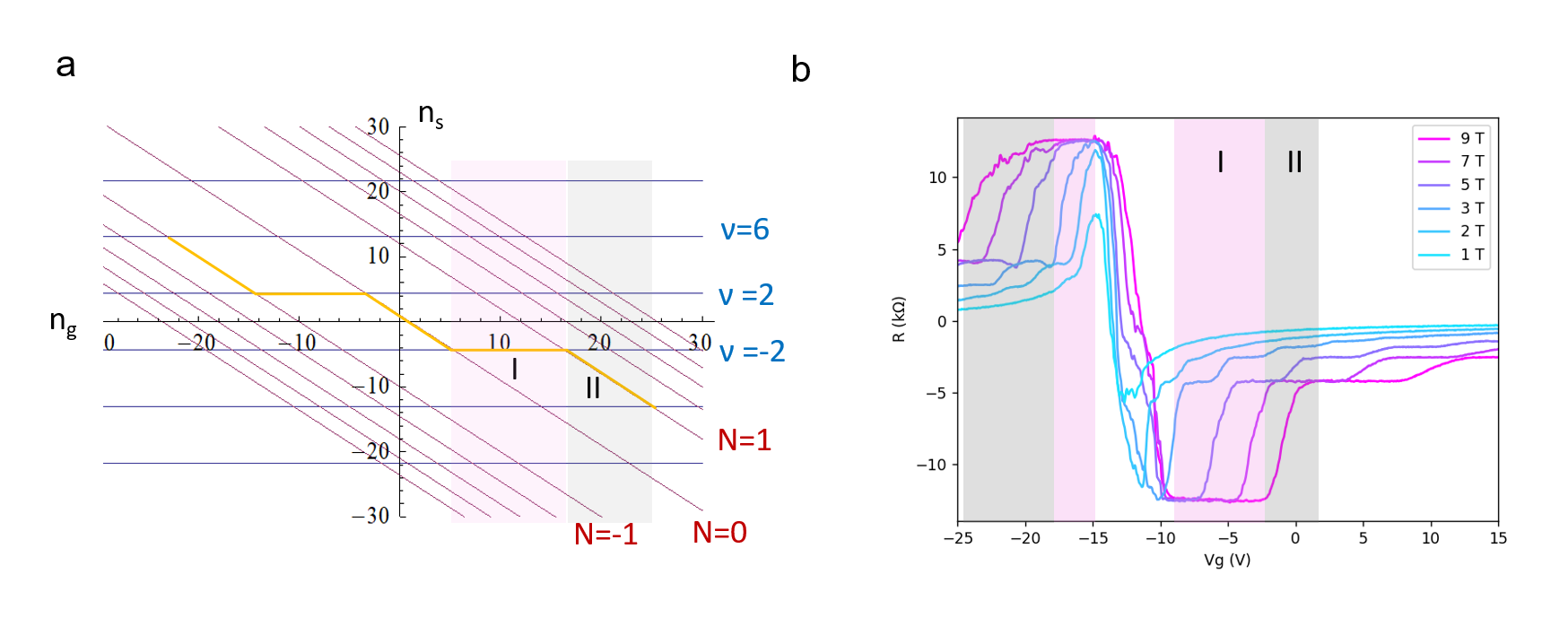} 

\caption{ a) Theoretical carrier density $n_s$ as a function of gate-induced charge density $n_g$ at B = 9 T. The blue line shows the linear relation derived from the quantum Hall condition $n_s = \nu B / \phi_0$, The red curve is based on the Landau level quantization model $\epsilon_F = v_F\sqrt{2\hbar e B |N|}$, capturing the density dependence through discrete energy levels indexed by integer $N$. b) Experimentally extracted carrier density versus gate voltage at high field from Hall resistance measurements.} 
\label{nsngB9}
\end{figure}

Under quantizing magnetic field, using the electrostatic gate to tune the carrier density is equivalent to varying the A-$\epsilon_F$ by $n_g(1/\gamma + e^2/C)$. The shift of Fermi level on the donor side is negligible compared to graphene as the density of states on the donor side is much larger. When the Fermi level crosses the interval between two Landau levels ( N=0, $\pm$1...) as shown in the region I of Figure \ref{nsngB9}, $n_s = \nu /\phi_0 B$, which does not depend on the $n_g$ and shown as the blue lines in  Figure \ref{nsngB9}a. While when the Fermi level crosses the Landau levels (region II), the Fermi energy in Eq. \ref{eq:cb} is notated as $\epsilon_F = v_F\sqrt{2\hbar e B |N|}$, resulting in the $n_s = \frac{A \gamma}{1+\gamma e^2/C} - \frac{ \gamma}{1+\gamma e^2/C} \sqrt{2 \hbar v_F ^2 e B |N|} - ng$. The width of Region I is $\Delta V_I = \frac{e}{C} \frac{ \gamma  v_F  }{1+\gamma e^2/C} \sqrt{2e\hbar B}$, and $\Delta V_{II} = \frac{e^2}{C} \frac{4B}{h}$. The ratio between these two regions can give information about the capacitance and the density of states at the interface, $w = \frac{\Delta V_{I}}{\Delta V_{II}} = \frac{ \gamma  v_F  }{1+\gamma e^2/C}  \frac {h^{3/2}}{4\sqrt{\pi e B} } $. At high field, the gate voltage width ratio between region I quantized v=2 plateau and region II v=2 to v=6 transition edge width depends on the C and $\gamma$. Take B=9T as an example, the w = 1.7 on the electron-doped side , giving the $\frac{ \gamma  }{1+\gamma e^2/C} = 1.5 \times 10^{13} eV^{-1} cm^{-2} $. We notice the width ratio is larger than the case of Graphene/BN stack, as well as the hole-doped side, which is likely from the large $\gamma$ density of states in the donor side. If we take the capacitance of $\rm SiO_2, 12.3 \ nF/ cm^2$, one will get $\gamma$ around $-7.7 \times 10 ^{10} \ eV^{-1} cm^{-2}$, where the negative sign likely means the hole accumulation on the $\rm PbI_2$ side.

\begin{figure}[h]

\includegraphics[clip=true,width=1.0\textwidth]{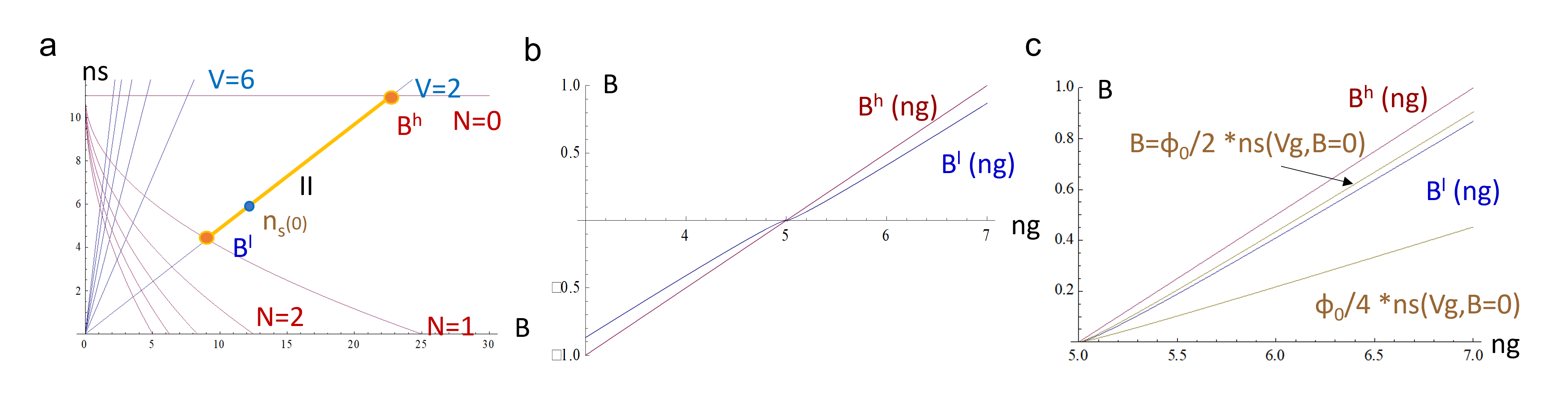} 

\caption{ $n_s$ as a function of $B$. a) from the theory model, the blue line is from $n_s = \nu B / \phi_0$, the red line is gotten from $\epsilon_F = v_F\sqrt{2\hbar e B |N|}$. The golden line indicates the v=2 plateau with the low boundary of $B^l$ and up boundary of $B^h$. The brown dot $n_s(0)$ refer to the calculated $\nu=2$ with zero-field density.  b) The trace of $B^l$ (blue) and $B^h$ (red) with ng. c) The B(ng) with the $\nu=2$  calculated from $n_s(0)$ (brown) by taking the ns(Vg) at low field (shown in Figure \ref{nsngB0}a) and $n_s$ = 2/$\phi_0$ B (upper branch), ns = 4/$\phi_0$ B (lower branch),  which is located between $B^l$ and  $B^h$.} 
\label{nsB}
\end{figure}

Under high field, at a fixed Vg, the field involves the  $n_s$ through $\epsilon_F$ quantized to $v_F \sqrt{2e\hbar B |N|}$  which crosses with $n_s = v \phi_0 B$ as depicted in part I. Here we select two points at two ends of $v=2$ plateau which are labeled in Figure \ref{nsB}a as $B^h$ (high boundary of $B(v=2)$) and $B^l$ (high boundary of $B(v=2)$). The solution itself is quite complicated to make numerical assumptions on A, $\gamma$, C, etc. so here we just follow their trends with $n_g$, as shown in  Figure \ref{nsB}b. The region between the $B^h(n_g)$ and $B^l(n_g)$ is expected to be the $\nu =2$ in the Landau fan diagram.   In  Figure \ref{nsB}c, we added the calculated trace of $\nu$=2 and $\nu$ = 4 with $n_s(B0)$ (gate dependent carriers density under zero field)

\vspace{5mm}

{\bf Disorders}
It has also been reported that high levels of disorder in graphene prevent the divergence at high magnetic fields, accompanied by large resistance fluctuations near the Dirac point, which was interpreted as a consequence of the presence of naturally distributed electron and hole puddles. \cite {martin2008observation,yang2016puddle,qiu2022mesoscopic} Such disordered systems typically exhibit low mobility and broad charge neutrality regions at low magnetic fields. In contrast, the mobility in our sample is good, and the behavior near the Dirac point deviates significantly from that of disorder-dominated systems. As shown in Fig. 1c and d in maintext, the longitudinal resistance $R_{xx}$ near the Dirac point decreases and eventually vanishes with increasing magnetic field. Concurrently, the Hall resistance $R_{xy}$ = 0 region shifts toward the electron-doped side, and a sharp resistance peak emerges at the Dirac point in Figure s3.  These behaviors are inconsistent with those of inhomogeneous systems governed by random disorder and instead suggest the presence of an underlying structured electronic phase. In addition, we observe coherent fluctuations on $R_{xx}$ at the quantized region from separated contact pairs, which we will discuss later. These observations indicate that if coexisting electron and hole regions are present in our device, they are not randomly distributed puddles but rather form a spatially ordered electronic structure. Such order could arise from periodic or quasi-periodic potentials, such as those induced by a moiré superlattice, enabling coherent conductive pathways and stabilizing long-range incompressible states.

\section{Moir\'e Hoffustadter spectrum}

\begin{figure}[h]

\includegraphics[clip=true,width=1\textwidth]{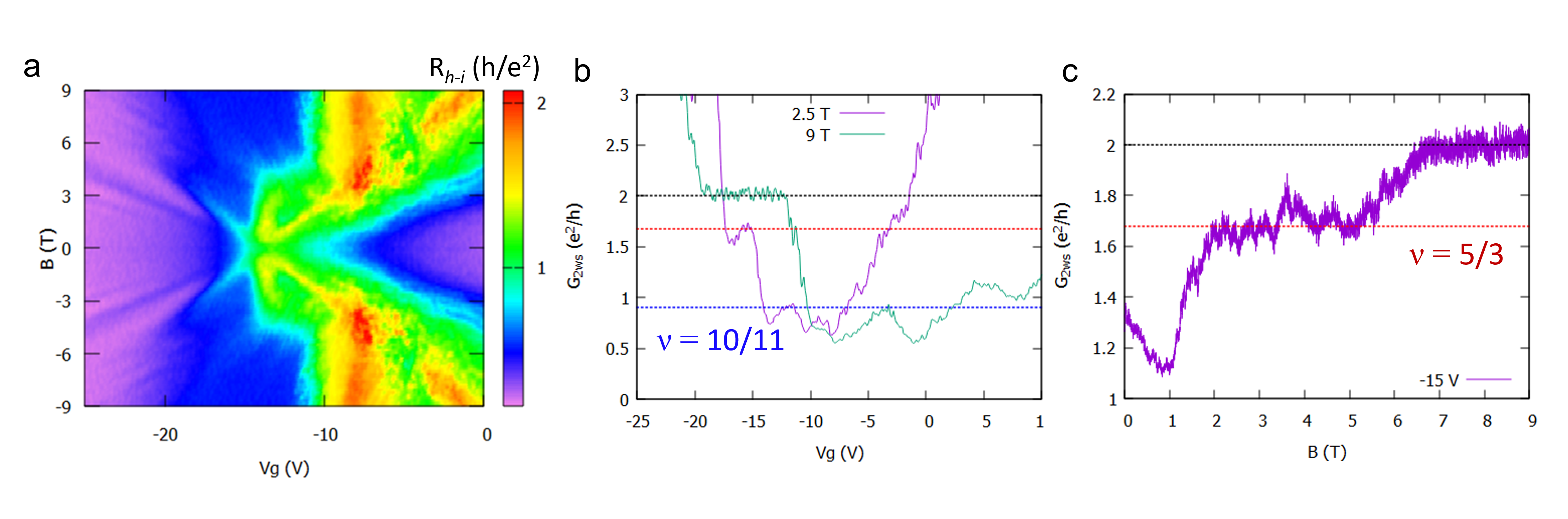} 

\caption{ a) Two-probe resistance as a function of magnetic field and gate voltage measured between contacts h-i. A pronounced high-resistance peak appears at carrier density $n_1 = 4.31 \times 10^{11}\ cm^{-2}$, corresponding to the first flavor of the moiré superlattice miniband.
  b) Two-probe conductance from the h-i contact pair as a function of gate voltage at 2.5 T and 9 T. At 2.5 T, two additional fractional conductance plateaus are identified at 10/11 $e^2/h$ (blue dashed line) and 5/3 $e^2/h$ (red dashed line), which can be attributed to possible 2/-10/6 and -2/10 Chern junctions, respectively. At 9 T, the quantized 2 $e^2/h$ plateau extends across the charge neutrality point and terminates near $n_1$ for B $>$ 1/4 $\phi_0$, implying that electronic transport is dominated by the moiré Hofstadter miniband.
  c) Two-probe conductance from contacts h-i as a function of magnetic field at Vg = -15 V, showing both quantized plateaus at 5/3 $e^2/h$ and 2 $e^2/h$.
} 
\label{hi}
\end{figure}

\begin{figure}[h]

\includegraphics[clip=true,width=0.8\textwidth]{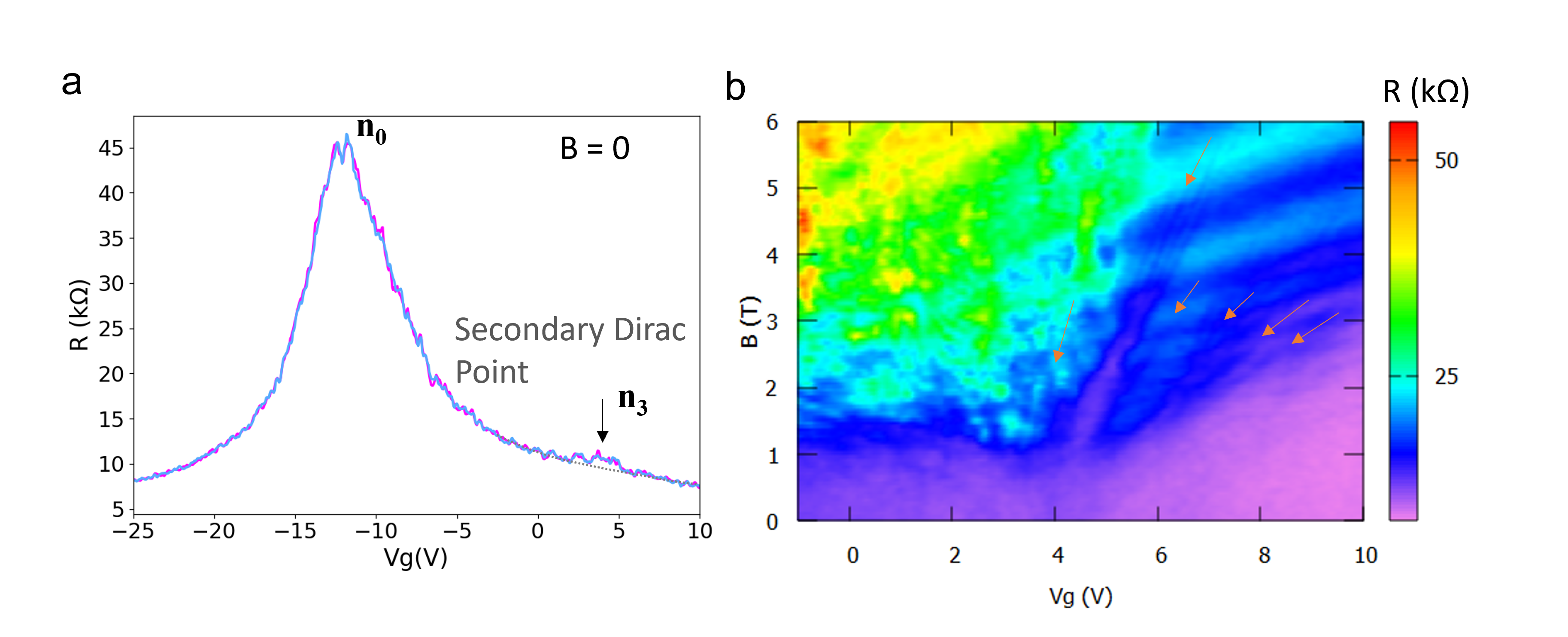} 

\caption{ a) Two-terminal resistance measured from the contact pair h–d as a function of gate voltage at zero magnetic field reveals a weak secondary Dirac point located at $n_3 = 12.93 \times 10^{11}\ \mathrm{cm}^{-2}$. b) Despite the weak zero-field feature, a prominent Landau fan emerges at intermediate magnetic fields. The arrows indicate linear trajectories corresponding to Chern numbers $\nu_m = 1, 2, 3, 4, 5, 6$. Notably, the $\nu_m = 2$ and $4$ trajectories each split into four finer lines at higher fields, suggesting the presence of additional miniband substructure or symmetry breaking.} 
\label{SDP}
\end{figure}

\begin{figure}[h]

\includegraphics[clip=true,width=0.5\textwidth]{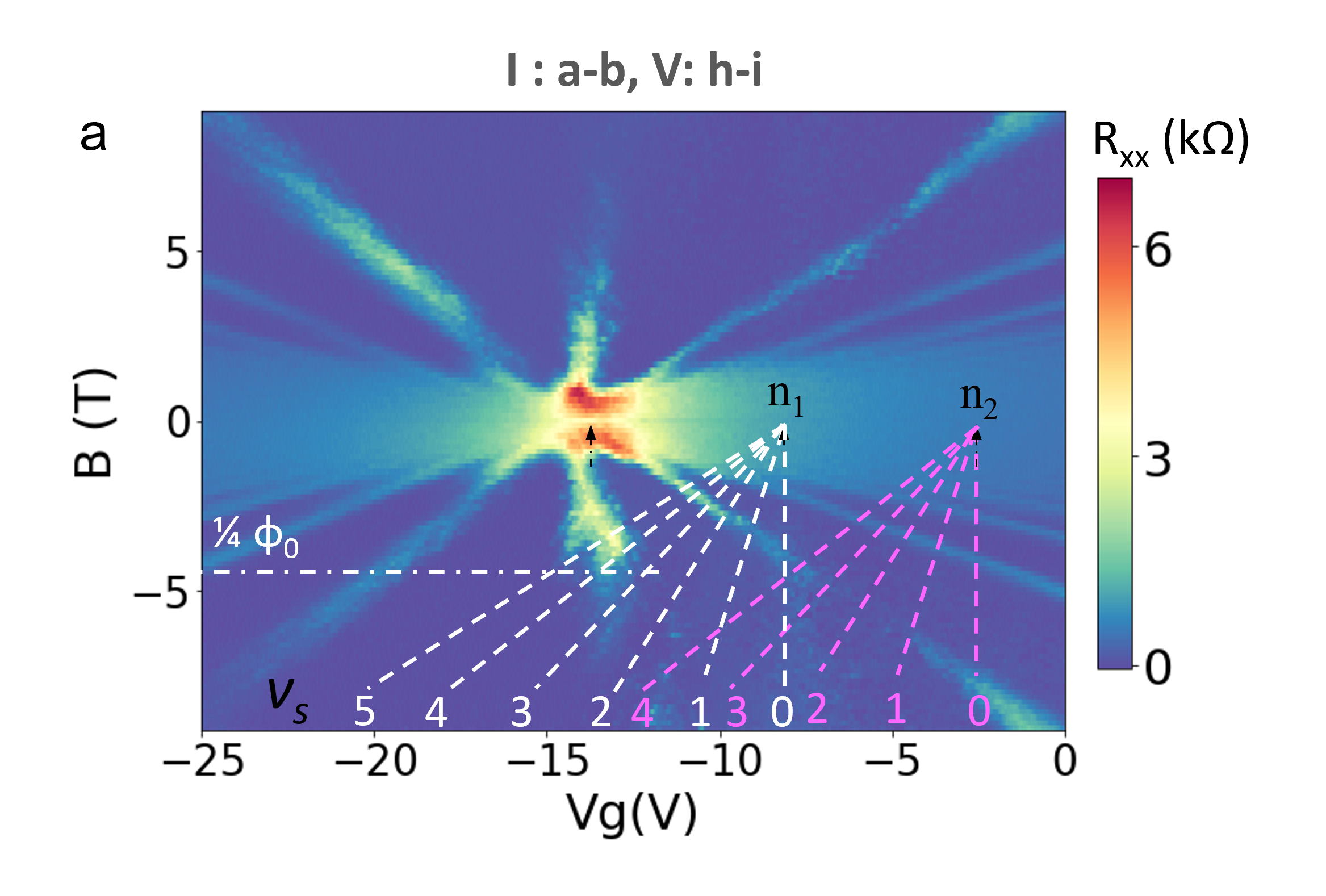} 

\caption{  The longitudinal resistance as a function of gate voltage and magnetic field from Fig. 1c in the main text. The linear trajectories follow the Diophantine equation with Chern numbers $\nu_m = 2,\ 3,\ 4,\ 5$ and moiré filling factor $s_m = 1$ (white dotted lines), coinciding with the boundaries of suppressed $R_{xx}$ at $\nu = 0$ (Dirac point). The purple dotted lines emanate from $s_m = 2$, with $\nu_m = 0,\ 1,\ 2,\ 3,\ 4,\ 5$, overlapping with vannishing resistance at the phase transition boundary between $\nu = 2$ and $\nu = 6$. The densities $\rm n_1 = 4.31 \times 10^{11}\ cm^{-2}$ and $\rm n_2 = 8.62 \times 10^{11}\ cm^{-2}$ match the values corresponding to the flavors shown in the two-terminal resistance. The onset magnetic field at the intersection is $\rm B = 1/4\ \phi_0$, which also marks the critical field for the emergence of the $2/3$ quantum Hall plateau. } 
\label{lineR0}
\end{figure}

\begin{figure}[h]

\includegraphics[clip=true,width=1.0\textwidth]{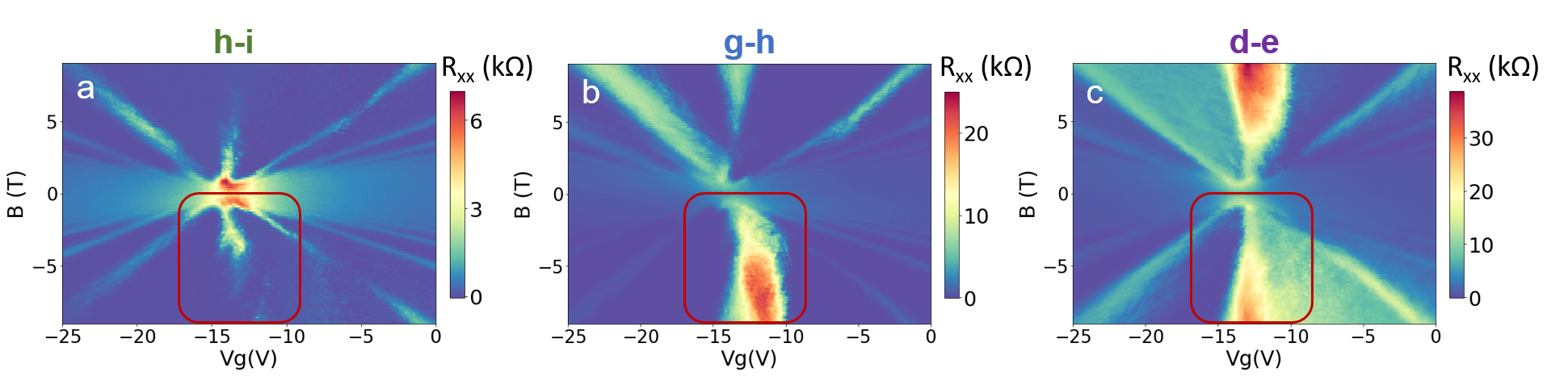} 

\caption{
Landau fan diagram of the longitudinal resistance measured from (a) h–i, (b) g–h, and (c) d–e contact pairs, with current sourced from contact a to b. The missing resistance peak at the charge neutrality point in the h–i configuration signals the presence of a moiré superlattice in that region, while finite resistance in the g–h and d–e configurations indicates that conventional quantum Hall transport dominates outside the moiré domain. The observed asymmetry between electron/hole doping and positive/negative magnetic fields likely arises from the positioning of contacts g and e near the edge of the $\mathrm{PbI}_2$/graphene heterostructure, where additional scattering and current leakage may occur.  } 
\label{Rxx3}
\end{figure}

\section{Spin orbit coupling}

\begin{figure}[h]

\includegraphics[clip=true,width=0.8\textwidth]{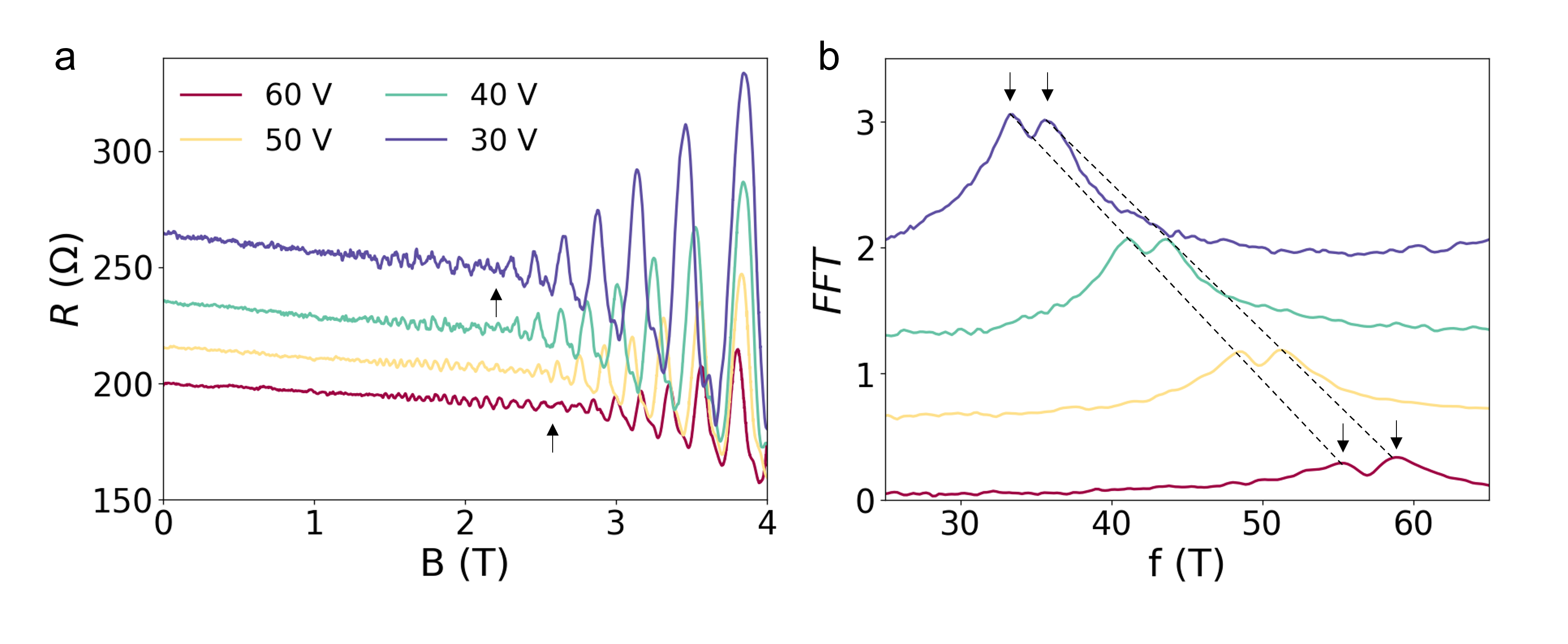} 

\caption{ The period of the Shubnikov–de Haas (SdH) oscillations provides a reliable method to estimate the size of the Fermi surface. Spin–orbit coupling (SOC) induces a splitting of the Fermi surface into subbands with opposite spin polarization in the electronic band structure. This splitting is manifested as a characteristic beating pattern in the SdH oscillations. Resolving the associated beating frequency offers a direct and unambiguous signature of Fermi surface splitting, thereby allowing for a quantitative measure of the spin splitting energy. Observation of such beating patterns at low magnetic fields requires exceptionally high device quality, particularly in terms of spatial homogeneity and carrier mobility.
In the graphene/$\mathrm{PbI}_2$ heterostructure, $\mathrm{PbI}_2$ induces spin–orbit interaction in graphene via the proximity effect. a) The SdH resistance oscillations exhibit a clear beating pattern, with node positions (marked by black arrows) shifting as a function of the applied gate voltage $V_g$. Specifically, the beating node moves to higher magnetic fields as $V_g$ increases. b) The corresponding Fourier spectra of the oscillations in (a) reveal a pair of peaks, with the frequency splitting increasing at higher gate voltages ( higher electron densities). These peak frequencies are proportional to the areas of the two spin-split Fermi surfaces. The observed splitting in the SdH oscillation frequencies directly reflects the SOC-induced Fermi surface splitting in graphene/$\mathrm{PbI}_2$.
 } 
\label{soc}
\end{figure}

\section{Reistance fluctuation}

\begin{figure}[h]

\includegraphics[clip=true,width=0.8\textwidth]{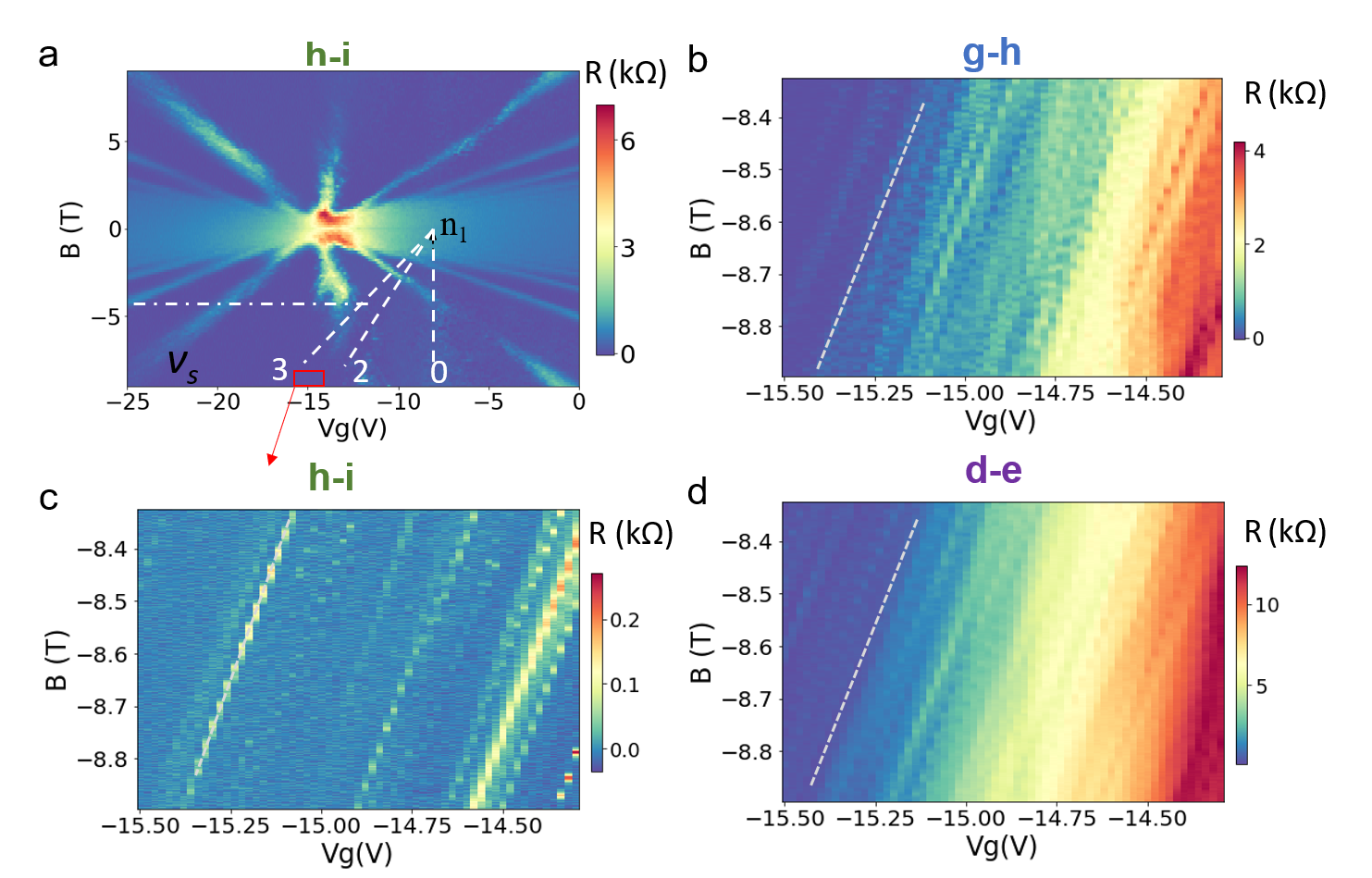} 

\caption{ The longitudinal resistance as a function of gate voltage and magnetic field exhibits linear fluctuation patterns associated with a Chern number $\nu_m = -2$. a) The Landau fan diagram from contact pair h-i with four-probe measurements, the same as the shown Fig. S11. We zoom into the hole-doped region with $\nu = -2$ near the Dirac point at $B > \phi_0/4$, where vanishing resistance is observed between contacts h–i (highlighted by the red box). The region also situates between $\nu_m = -2$ and $\nu_m = -3$. b–d) Magnified resistance maps from contact pairs g–h, h–i, and d–e, respectively, show distinct fluctuation features that align with Diophantine trajectories characterized by $\nu_m = -2$. Notably, the resistance fluctuations measured from g–h and d–e exhibit magnetic-field-synchronized behavior over selected intervals, as discussed in Figure 4 of the main text. } 
\label{Rxxfluctuation}
\end{figure}

\begin{figure}[h]

\includegraphics[clip=true,width=0.5\textwidth]{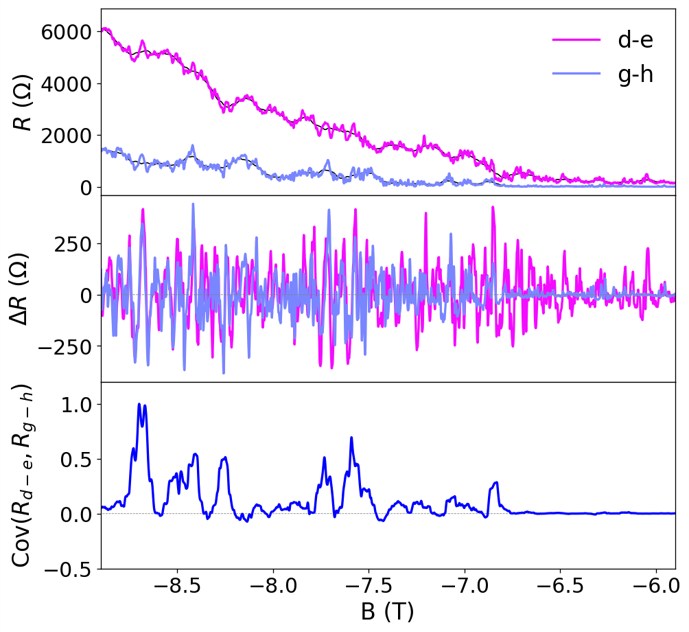} 

\caption{ The correlated resistance fluctuations measured from contact pairs d–e and g–h in the magnetic field range from –5.9 T to –8.9 T exhibit minimal correlation below –6.7 T, but reveal obvious correlation at higher magnetic fields.  } 
\label{longrangecov}
\end{figure}

\end{document}